\documentclass[aps, prd, letterpaper, twocolumn, amsmath, amssymb, superscriptaddress, floatfix, nofootinbib, longbibliography]{revtex4-1}
\usepackage[T1]{fontenc}
\usepackage[english]{babel}
\usepackage[utf8]{inputenc}
\usepackage{graphicx}
\usepackage{color}
\usepackage{xcolor}
\usepackage{mathrsfs}
\usepackage{amsfonts}
\usepackage[colorlinks=true, linkcolor=blue, urlcolor=blue, citecolor=blue, anchorcolor=blue]{hyperref}

\newcommand{\cD}{\ensuremath{\mathcal D} }
\newcommand{\scrD}{\ensuremath{\mathscr D} }
\newcommand{\Ibb}{\ensuremath{\mathbb I} }
\newcommand{\cM}{\ensuremath{\mathcal M} }
\newcommand{\cMdag}{\ensuremath{\mathcal M^{\dag}} }
\newcommand{\cN}{\ensuremath{\mathcal N} }
\newcommand{\cO}{\ensuremath{\mathcal O} }
\newcommand{\Pbb}{\ensuremath{\mathbb P} }
\newcommand{\al}{\ensuremath{\alpha} }
\newcommand{\be}{\ensuremath{\beta} }
\newcommand{\ga}{\ensuremath{\gamma} }

\newcommand{\de}{\ensuremath{\delta} }
\newcommand{\De}{\ensuremath{\Delta} }
\newcommand{\chimax}{\ensuremath{\chi_{\text{max}}} }
\newcommand{\eps}{\ensuremath{\epsilon} }
\newcommand{\la}{\ensuremath{\lambda} }
\newcommand{\lalat}{\ensuremath{\la_{\text{lat}}} }
\newcommand{\mulat}{\ensuremath{\mu_{\text{lat}}} }
\newcommand{\si}{\ensuremath{\sigma} }
\newcommand{\nn}{\nonumber }
\newcommand{\gsim}{\ensuremath{\gtrsim} }
\newcommand{\lsim}{\ensuremath{\lesssim} }
\newcommand{\SO}[1]{\ensuremath{\text{SO(}#1\text{)}}}
\newcommand{\X}{\ensuremath{\!\times\!} }
\newcommand{\chidof}{\ensuremath{\mbox{$\chi^2/\text{d.o.f.}$}} }
\newcommand{\pf}{\ensuremath{\text{pf}\,} }
\newcommand{\Tr}[1]{\ensuremath{\text{Tr}\left[ #1 \right]} }
\newcommand{\vev}[1]{\ensuremath{\left\langle #1 \right\rangle} }
\newcommand{\eq}[1]{Eq.~(\ref{#1})}
\newcommand{\fig}[1]{Fig.~\ref{#1}}
\newcommand{\tab}[1]{Table~\ref{#1}}
\newcommand{\secref}[1]{Sec.~\ref{#1}}
\newcommand{\appref}[1]{App.~\ref{#1}}
\newcommand{\refcite}[1]{Ref.~\cite{#1}}
\newcommand{\crit}{\ensuremath{\left.\frac{T}{\mu}\right|_{\text{crit.}}} } 
\newcommand{\critinline}{\ensuremath{\left(T / \mu\right)_{\text{crit.}}} }

\begin{document}
\title{Finite-temperature phase diagram of the BMN matrix model on the lattice}

\author{Raghav G. Jha}\email{raghav.govind.jha@gmail.com}
\affiliation{Thomas Jefferson National Accelerator Facility, Newport News, VA 23606, United States}

\author{Anosh~Joseph}\email{anosh.joseph@wits.ac.za}
\affiliation{National Institute for Theoretical and Computational Sciences, School of Physics, and Mandelstam Institute for Theoretical Physics, University of the Witwatersrand, Johannesburg, Wits 2050, South Africa}

\author{David Schaich}\email{david.schaich@liverpool.ac.uk}
\affiliation{Department of Mathematical Sciences, University of Liverpool, Liverpool L69 7ZL, United Kingdom}

\begin{abstract}
We investigate the thermal phase structure of the Berenstein--Maldacena--Nastase (BMN) matrix model using non-perturbative lattice Monte Carlo calculations.
Our main analyses span three orders of magnitude in the coupling, involving systems with sizes up to $N_{\tau} = 24$ lattice sites and SU($N$) gauge groups with $8 \leq N \leq 16$.
In addition, we carry out extended checks of discretization artifacts for $N_{\tau} \leq 128$ and gauge group SU(4).
We find results for the deconfinement temperature that interpolate between the perturbative prediction at weak coupling and the large-$N$ dual supergravity calculation at strong coupling.
While we confirm that the phase transition is first order for strong coupling, it appears to be continuous for weaker couplings.
\end{abstract}

\maketitle

\section{Introduction}
\label{sec:intro}

The idea of gauge/gravity duality allows us to investigate aspects of quantum gravity by studying the dual gauge theory and vice versa.
For example, the thermodynamic properties of maximally supersymmetric Yang--Mills (SYM) theories in $(p + 1)$ dimensions with large SU($N$) gauge groups and strong 't~Hooft couplings capture the semi-classical thermodynamics of non-extremal black D$p$ branes~\cite{Itzhaki:1998dd}.
The first concrete gauge/gravity duality proposal~\cite{Maldacena:1997re} considered $p = 3$, conjecturing an equivalence between conformal $\cN = 4$ SYM and Type~IIB supergravity on AdS$_5 \times S^5$.
In lower dimensions $p < 3$, the situation is more complicated because the supersymmetric gauge theories are not conformal.
This motivates the use of non-perturbative lattice field theory calculations to analyze these strongly coupled gauge theories to gain insights into quantum aspects of the dual gravitational systems.

In recent years, there has been progress in extracting $p < 2$ black hole physics from numerical Monte Carlo analyses of the dual strongly coupled gauge theories at finite temperatures and large $N$~\cite{Catterall:2007fp, Anagnostopoulos:2007fw, Catterall:2008yz, Catterall:2008dv, Hanada:2008gy, Hanada:2008ez, Catterall:2009xn, Nishimura:2009xm, Catterall:2010gf, Catterall:2010fx, Catterall:2011aa, Kadoh:2012bg, Hanada:2013rga, Giguere:2015cga, Kadoh:2015mka, Filev:2015hia, Hanada:2016zxj, Berkowitz:2016tyy, Berkowitz:2016jlq, Kadoh:2017mcj, Rinaldi:2017mjl, Catterall:2017lub, Jha:2017zad, Berkowitz:2018qhn, Asano:2018nol, Schaich:2020ubh, Bergner:2021goh, Schaich:2022duk, Pateloudis:2022oos, Pateloudis:2022ijr}, which is currently being extended to $p = 2$~\cite{Catterall:2020nmn, Sherletov:2022rnl, Sherletov:2023udh, Schaich:2024LAT}.
See \refcite{Schaich:2022xgy} for a recent review of these developments, including further references.
In the context of lattice calculations, the lower dimensionality helps both to reduce the computational costs as well as to simplify the process of taking the continuum limit: As discussed in Refs.~\cite{Golterman:1988ta, Giedt:2004vb, Bergner:2007pu, Catterall:2007fp, Giedt:2018ygt}, little to no fine-tuning is required to recover the supersymmetric target theory in the continuum limit.

When $p = 0$, maximal SYM reduces to a quantum-mechanical theory~\cite{deWit:1988wri} best known due to a conjecture by Banks, Fischler, Shenker and Susskind (BFSS)~\cite{Banks:1996vh} that the large-$N$ limit of this model describes the strong-coupling (M-theory) limit of Type~IIA string theory in the infinite-momentum frame.
Several groups have numerically studied this `BFSS model', using either a lattice discretization of (Euclidean) time~\cite{Catterall:2007fp, Catterall:2008yz, Catterall:2009xn, Kadoh:2012bg, Kadoh:2015mka, Filev:2015hia, Berkowitz:2016tyy, Berkowitz:2016jlq, Rinaldi:2017mjl, Berkowitz:2018qhn, Pateloudis:2022ijr} or a `non-lattice' momentum-space approach~\cite{Anagnostopoulos:2007fw, Hanada:2008gy, Hanada:2008ez, Nishimura:2009xm, Hanada:2013rga, Hanada:2016zxj}.
These investigations have reached sufficiently low temperatures and large $N$ to provide confidence that the observed thermodynamic behavior approaches the leading-order prediction of the dual Type~IIA supergravity.
Fits to these numerical Monte Carlo results have also been used to predict $\al'$ and $g_s$ corrections to the classical supergravity solutions from the dual SYM quantum mechanics~\cite{Hanada:2008ez, Kadoh:2015mka, Hanada:2016zxj, Berkowitz:2016tyy, Berkowitz:2016jlq, Pateloudis:2022ijr}.

In the BFSS model, the thermal partition function is not well defined since it includes integration over a non-compact moduli space.
This instability cannot be seen directly in the black D0-brane thermodynamics described by large-$N$ supergravity, as it is a $1/N$ effect.
Even so, it means that Monte Carlo calculations of the BFSS model are unstable for any finite $N$~\cite{Catterall:2009xn}; as $N$ increases, the system may spend longer fluctuating around a metastable vacuum, but this will eventually decay given sufficient time.
The standard way to address this issue and stabilize numerical calculations is to deform the theory by adding a regulator that gives a small mass to the scalars.
This explicitly breaks supersymmetry, and the extrapolation necessary to remove this deformation can significantly increase the complexity and costs of the calculations.

This motivates the alternative of studying the Berenstein--Maldacena--Nastase (BMN) deformation of the BFSS model~\cite{Berenstein:2002jq}, which describes the discrete light-cone quantization compactification of M-theory on the maximally supersymmetric ``pp-wave'' background of 11d supergravity.
The `BMN model' preserves maximal supersymmetry and has a well-defined thermal partition function because of the absence of flat directions, making it a better-behaved system to study using Monte Carlo methods.
%
Among its many notable features~\cite{Horowitz:1989bv, Kim:2003rza, Lin:2005nh, Ishii:2008ib, Ishiki:2009sg, Ishiki:2011ct, Costa:2014wya}, the BMN model possesses a non-trivial finite-temperature phase diagram with a high-temperature deconfined phase and a low-temperature confined phase.\footnote{While the BFSS model has historically been expected to be deconfined for all temperatures, Refs.~\cite{Bergner:2021goh, Pateloudis:2022ijr} recently argued that a confined phase persists in the BFSS limit in which the BMN deformation is removed.}
In this sense, it resembles higher-dimensional theories~\cite{Catterall:2017lub, Jha:2017zad, Catterall:2020nmn, Sherletov:2022rnl, Sherletov:2023udh, Schaich:2024LAT}, while involving more modest computational costs in numerical analyses.
However, the gravity dual of the BMN model is more complicated than that of the BFSS model and, therefore, is less explored~\cite{Lin:2005nh, Costa:2014wya}.

In this work, we study the phase diagram of the BMN model on the lattice, determining critical deconfinement transition temperatures at a range of couplings spanning three orders of magnitude from the perturbative regime to stronger couplings that approach the supergravity solution.
We carry out calculations with multiple lattice sizes and numbers of colors $N$ while also checking discretization artifacts for our lattice action and evaluating the Pfaffian of the fermion operator to ensure we do not encounter a sign problem.
This finalizes the investigations reported in preliminary form by our recent conference proceedings~\cite{Schaich:2020ubh, Schaich:2022duk}.
We begin in the next section by reviewing the BMN model and previous lattice studies of it.
In \secref{sec:latform}, we discuss the simple lattice formulation that we employ in this work, also addressing discretization artifacts and the Pfaffian.
Our numerical results for the phase diagram are presented in \secref{sec:results}, including comparisons to predictions from perturbation theory and the dual supergravity, as well as analyses of the order of the transition.
The data leading to these results are available through \refcite{data}.
We conclude in \secref{sec:conc} with a discussion of our plans for future work.

\section{BMN Matrix Model}
\label{sec:BMN_Matrix_Model}

Our starting point is the BFSS model in Euclidean time $\tau$.
With an anti-Hermitian basis for the SU($N$) generators, $\Tr{T^a T^b} = - \de^{ab}$, the continuum action takes the form
\begin{align}
  S_{\text{BFSS}} = \frac{N}{4 \la} \int d\tau \ \mbox{Tr} \Bigg[ & -\left(D_{\tau} X_i \right)^2 - \frac{1}{2} \sum_{i < j} \left[ X_i, X_j \right]^2 \cr
                                                                  & + \Psi_{\al}^T \ga_{\tau}^{\al \be} D_{\tau} \Psi_{\be} \label{eq:bfss} \\
                                                                  & + \frac{1}{\sqrt{2}} \Psi_{\al}^T \ga_i^{\al \be} \left[X_i, \Psi_{\be} \right] \Bigg], \nn
\end{align}
where $D_{\tau} = \partial_{\tau} + [A, ~\cdot~]$ is the covariant derivative corresponding to the gauge field $A$, $X_i$ are the nine scalars of the theory, and $\Psi_{\al}$ is a sixteen-component spinor.
The indices $i, j = 1, \cdots, 9$ while $\al, \be = 1, \cdots, 16$; the latter will generally be suppressed, and we have also suppressed the $\tau$ dependence of all the fields.
$\ga_i$ and $\ga_{\tau}$ are $16 \X 16$ Euclidean gamma matrices; in the next section, we will specify the representation we employ for them. 
All fields transform in the adjoint representation of the SU($N$) gauge group.
In this ($0+1$)-dimensional setting, both the 't~Hooft coupling $\la \equiv g_{\text{YM}}^2 N$ and the Yang--Mills coupling $g_{\text{YM}}^2$ are dimensionful.

The theory defined by \eq{eq:bfss} has flat directions when the scalar fields $X_i$ commute with each other.
Supersymmetry preserves some of these flat directions at the quantum level, which results in an ill-defined thermal partition function as discussed in \cite{Catterall:2009xn}.
The BMN model changes this by adding the following terms to the action:
\begin{align}
  S_{\mu} = - \frac{N}{4 \la} \int d\tau \ \mbox{Tr} \Bigg[ & \left( \frac{\mu}{3} X_I \right)^2 + \left( \frac{\mu}{6} X_M \right)^2   \cr
                                                            & + \frac{\mu}{4} \Psi_\al^T \ga_{123}^{\al \be} \Psi_\be \label{eq:deform} \\
                                                            & + \frac{\sqrt{2} \mu}{3} \eps_{IJK} X_I X_J X_K \Bigg], \nn
\end{align}
with dimensionful deformation parameter $\mu$.
The indices $i, j$ divide into the two sets where $I, J, K$ take values in $1, 2, 3$ while $M = 4, \cdots, 9$.
We also define
\begin{equation}
  \ga_{123} \equiv \frac{1}{3!} \eps_{IJK} \ga_I \ga_J \ga_K = \ga_1 \ga_2 \ga_3.
\end{equation}
For non-zero $\mu$, the terms in \eq{eq:deform} explicitly break the SO(9) global symmetry of \eq{eq:bfss} down to $\SO{6} \X \SO{3}$ while preserving all 16 supersymmetries of the theory.
Integrating over the fermions produces the Pfaffian of the fermion operator, $\pf(\cM)$.
The Euclidean path integral in the continuum then takes the form
\begin{equation}
  Z = \int [\scrD A] [\scrD X] \ \pf(\cM) \ e^{-S_B},
\end{equation}
where the bosonic action is
\begin{align}
  S_B = - \frac{N}{4 \la} \int d\tau \ \mbox{Tr} \Bigg[ & \left( D_{\tau} X_i \right)^2 + \frac{1}{2} \sum_{i < j} \left[ X_i, X_j \right]^2    \cr
                                                      & + \left( \frac{\mu}{3} X_I \right)^2 + \left( \frac{\mu}{6} X_M \right)^2 \\
                                                      & + \frac{\sqrt{2} \mu}{3} \eps_{IJK} X_I X_J X_K \Bigg], \nn
\end{align}
and the fermion operator is
\begin{equation}
  \label{eq:fermion_op}
  \cM(A, X) = \ga_{\tau} D_{\tau} + \ga_i [X_i, ~\cdot~] - \frac{\mu}{4} \ga_{123}.
\end{equation}

In addition to the usual classical vacuum with $X_i = 0$, this theory also has a large number of `fuzzy-sphere' vacua in which the \SO{3} scalars take non-zero values that satisfy the relation
\begin{equation}
  \label{eq:fuzzy}
  X_I \propto \epsilon_{IJK} X_J X_K.
\end{equation}
These vacua do not play a role in our current numerical investigations, which can be confirmed through our open data release~\cite{data}.
In particular, as we will discuss further below, we monitor several quantities that are sensitive to fuzzy-sphere contributions, including the gauge-invariant `scalar squares' $\Tr{X_i^2}$ in addition to the scalar-trilinear `Myers' term
\begin{equation}
  \label{eq:Myers}
  \widehat{M} \equiv \frac{M}{\la} = \frac{\sqrt{2}}{12 N} \frac{1}{\la \be} \vev{\int d\tau \ \eps_{IJK} \ \mbox{Tr} \left(X_I X_J X_K \right)},
\end{equation}
which we make dimensionless by including a factor of $\la$.

In \eq{eq:Myers} we have introduced a finite temperature $T$ by formulating the BMN model on a temporal circle of circumference $\be = 1 / T$.
The non-zero temperature breaks supersymmetry, and according to gauge/gravity duality, the finite-temperature theory corresponds to a black-hole geometry in the dual supergravity.
To make sure that the $\al'$ corrections near the horizon of the dual black hole are small, the system must be in the regime $T / \la^{1/3} \ll 1$ and $\mu / \la^{1/3} \ll 1$.
It is convenient to define the dimensionless coupling $g \equiv \la / \mu^3$, in terms of which the latter constraint is $g \gg 1$.
In our numerical calculations, we will identify the critical temperature of the deconfinement transition by fixing $g$ and scanning in the dimensionless ratio $T / \mu$, which we will call just ``the temperature''.

There are predictions for the critical temperature \critinline in both the weak- and strong-coupling limits.
In weak-coupling limit $g \to 0$, perturbative calculations~\cite{Furuuchi:2003sy, Spradlin:2004sx, Hadizadeh:2004bf} predict a first-order deconfinement transition with critical temperature
\begin{equation}
  \label{eq:weak}
  \lim_{g \to 0} \crit = \frac{1}{12\ln 3} \approx 0.076.
\end{equation}
In the $N \to \infty$ planar limit, Refs.~\cite{Spradlin:2004sx, Hadizadeh:2004bf} also calculate the $\cO(g)$ and $\cO(g^2)$ corrections,
\begin{equation}
  \label{eq:NNLO}
  \crit = \frac{1}{12\ln 3} \left[1 + C_{\text{NLO}} \left(27 g\right) - C_{\text{NNLO}}\left(27 g\right)^2 + \cdots\right]
\end{equation}
where the perturbative expansion parameter is $27g$, while
\begin{align*}
              C_{\text{NLO}} & = \frac{2^6 \cdot 5}{3^4} \approx 4 \cr
              \text{and~} C_{\text{NNLO}} & = \frac{23\cdot 19\,927}{2^2\cdot 3^7} + \frac{1\,765\,769\ln 3}{2^4\cdot 3^8} \approx 71.
\end{align*}
For strong coupling $g \to \infty$ with $T \ll \la^{1/3}$, the numerical construction of the dual supergravity solutions at finite temperature by \refcite{Costa:2014wya} predicts a larger
\begin{equation}
  \label{eq:strong}
  \lim_{g \to \infty} \crit \approx 0.106,
\end{equation}
with the transition still first order.

Our ambition is to use non-perturbative lattice calculations to track \critinline as the system interpolates between the weak-coupling limit of \eq{eq:weak} and the strong-coupling limit of \eq{eq:strong}, similar to what we did for the bosonic sector of the BMN model in Refs.~\cite{Dhindsa:2022uqn, Dhindsa:2023oes}.
This goal is strongly motivated by the duality with quantum gravity discussed above, as well as by the small number of lattice investigations so far.
Only three other groups have previously studied the phase diagram of the BMN model on the lattice~\cite{Catterall:2010gf, Asano:2018nol, Bergner:2021goh}.\footnote{More recent lattice studies of the BMN model have focused on `ungauging'~\cite{Pateloudis:2022oos} or the energy of the dual black D$0$-branes~\cite{Pateloudis:2022ijr} as opposed to the phase diagram of interest here.}
All of these three previous lattice studies take different approaches to explore the phase diagram of the theory, which also differ from our approach presented below.

The earliest study, \refcite{Catterall:2010gf}, was limited to a small number of lattice sites ($N_{\tau} = 5$) and small numbers of colors ($N = 3$ and 5).
Apart from constant rescalings of fields and parameters, this work uses the same lattice action as we describe in the next section.
Converting to our conventions, \refcite{Catterall:2010gf} fixes $T / \mu = 1 / 3$ and scans in the coupling $g$ --- essentially the opposite of our approach --- finding a deconfinement transition with critical coupling $g_{\text{crit.}} \simeq 0.035$. 
While $T / \mu = 1 / 3$ is larger than both the $g \to 0$ and $g \to \infty$ limits discussed above, the small values of $N_{\tau}$ and $N$ considered by \refcite{Catterall:2010gf} may make those limits inapplicable.

In addition to significantly increasing the number of lattice sites (focusing on $N_{\tau} = 24$) and colors (focusing on $N = 8$), \refcite{Asano:2018nol} also employs a second-order discretization of the covariant derivative, which introduces four new parameters whose values were chosen to minimize lattice artifacts in the inverse Dirac operator.
This work fixes $g$ by considering fixed values of $\mu / \la^{1 / 3} = 1 / g^{1 / 3}$, focusing on $9 \geq \mu / \la^{1 / 3} \geq 1$ that correspond to $0.00137 \leq g \leq 1$.
For each fixed $\mu / \la^{1 / 3}$, \refcite{Asano:2018nol} scans in a dimensionless temperature $T / \la^{1 / 3}$, which also differs from our approach described below.
For larger $\mu / \la^{1 / 3} \gsim 3$ (weaker couplings $g \lsim 0.037$), this work observes distinct deconfinement and SO(9)-symmetry-breaking transitions, all first order, which merge for larger $g$. 

Finally, \refcite{Bergner:2021goh} employs gauge fixing to the static diagonal gauge, $A(\tau) = \mbox{diag}(\al_1, \cdots, \al_N) / \be$, which reduces the number of degrees of freedom and allows investigations of lattice sizes up to $N_{\tau} = 48$ and up to $N = 48$ colors.
This work again fixes $\mu / \la^{1/3}$ and scans in $T / \la^{1 / 3}$, focusing on smaller $5/3 \geq \mu / \la^{1 / 3} \leq 0.1$ that imply stronger couplings $0.216 \leq g \leq 1000$, with particular interest in the $\mu \to 0$ limit where the BMN model reduces to the BFSS model. 
\refcite{Bergner:2021goh} confirms the first-order nature of the transition by presenting two-state signals in the Polyakov loop for $2/3 \gsim \mu / \la^{1/3} \geq 0.17$ corresponding to $3.4 \lsim g \leq 216$. 
For $\mu / \la^{1 / 3} \approx 1 / 3 \to g \approx 27$, the resulting \critinline agrees with the strong-coupling limit in \eq{eq:strong}.
Larger critical temperatures at stronger couplings, with fixed $N = 12$, are attributed to finite-$N$ corrections.

\section{Lattice Formulation and discretization artifacts}
\label{sec:latform}

Like \refcite{Catterall:2010gf}, and unlike Refs.~\cite{Asano:2018nol, Bergner:2021goh}, we use a simple gauge-invariant lattice discretization to formulate the BMN matrix model on a one-dimensional Euclidean lattice with thermal boundary conditions (periodic for the bosons and anti-periodic for the fermions).
The temporal extent of the lattice, $\be = a N_{\tau}$, is divided among $N_{\tau}$ sites separated by lattice spacing `$a$'.
The corresponding temperature is $T = 1 / (a N_{\tau})$.
Our numerical calculations involve the dimensionless lattice parameters $\mulat = a \mu$ and $\lalat = a^3 \la$.
Rather than considering the ratio $T / \la^{1 / 3}$ used by Refs.~\cite{Asano:2018nol, Bergner:2021goh}, we focus on the dimensionless parameters introduced in the previous section, which remain consistent in both the lattice and continuum theories:
\begin{align}
  \label{eq:params}
  \frac{T}{\mu} & = \frac{1}{N_{\tau} \mulat} &
  g & \equiv \frac{\la}{\mu^3} = \frac{\lalat}{\mulat^3}.
\end{align}
In a similar way, because we made the Myers term dimensionless in \eq{eq:Myers}, on the lattice it is simply
\begin{equation}
  \label{eq:Myers_lat}
  \widehat{M} = \frac{\sqrt{2}}{12 N \lalat N_{\tau}} \vev{\sum_{n = 0}^{N_{\tau} - 1} \eps_{IJK} \ \mbox{Tr} \left(X_I X_J X_K \right)}.
\end{equation}

We use the following lattice discretization of the covariant derivative:
\begin{equation}
  \cD_{\tau} = \begin{pmatrix}0            & \cD_{\tau}^+ \\
                              \cD_{\tau}^- & 0\end{pmatrix},
\end{equation}
where $\cD_{\tau}^-$ is the adjoint of $\cD_{\tau}^+$.
The finite-difference operator $\cD_{\tau}^+$ acts on the fermions at lattice site $n$ as
\begin{equation}
  \cD_{\tau}^+ \Psi_n = U(n) \Psi_{n + 1} U^{\dag}(n) - \Psi_n,
\end{equation}
where $U(n)$ is the Wilson gauge link connecting site $n$ and site $n + 1$.
That is, under an SU($N$) lattice gauge transformation, $U(n) \to G(n) U(n) G^{\dag}(n + 1)$.
$U^{\dag}(n)$ is the adjoint link with the opposite orientation.
Our choice of $\cD_{\tau}$ provides the correct number of fermions, free from extraneous `doublers', which is readily seen by setting the gauge links to unit matrices and computing $\det \cD_{\tau} = \det \left(\De^+ \De^-\right)$, the determinant of the scalar Laplacian.

The resulting bosonic action of the lattice theory is
\begin{align}
  S_B = - \frac{N}{4 \lalat} \sum_{n = 0}^{N_{\tau} - 1} \mbox{Tr} & \Bigg[ \left( \cD_{\tau} X_n^i \right)^2 + \frac{1}{2} \sum_{i < j} \left[ X_n^i, X_n^j \right]^2 \nn \\
                                                                   & + \left( \frac{\mulat}{3} X_n^I \right)^2 + \left( \frac{\mulat}{6} X_n^M \right)^2  \\
                                                                   & + \frac{\sqrt{2} \mulat}{3} \eps_{IJK} X_n^I X_n^J X_n^K \Bigg], \nn
\end{align}
while the fermion operator $\cM(U, X)$ matches \eq{eq:fermion_op} with the covariant derivative $D_{\tau}$ replaced by the finite-difference operator $\cD_{\tau}$.
For the gamma matrices we use a representation where $\ga_8$ and $\ga_9 = -i\Ibb_{16}$ are diagonal while
\begin{equation}
  \label{eq:gamma10}
  \ga_{\tau} = \begin{pmatrix}0 & \Ibb_8 \\ \Ibb_8 & 0\end{pmatrix} = \si_1 \otimes \Ibb_8
\end{equation}
and the other \ga matrices are all real, with $i\si_2$ in the outermost Kronecker product --- see \appref{app:gamma} for further details or \refcite{susy_code} for the explicit construction.

The lattice action as a whole is finite in lattice perturbation theory.
The only divergences that can occur arise from a one-loop fermion tadpole, which vanishes in a non-abelian theory because of gauge invariance.
This is shown for the BFSS model in \refcite{Catterall:2007fp}, which remains applicable in the BMN case.
Hence, at the classical level, the system will flow to the correct continuum supersymmetric target theory without fine-tuning as the lattice spacing is reduced.
Thus, the continuum limit is obtained by extrapolating $N_{\tau} \to \infty$ with $g$ and $T / \mu$ fixed, while the thermodynamic limit corresponds to increasing the number of colors, $N \to \infty$.

In numerical calculations, we use the standard rational hybrid Monte Carlo (RHMC) algorithm~\cite{Clark:2006fx}, which we have implemented in the publicly available parallel software package for lattice supersymmetry~\cite{susy_code} presented by \refcite{Schaich:2014pda}.
In contrast to the higher-dimensional theories that \refcite{Schaich:2014pda} focuses on, for the BMN model, we employ the gauge-invariant Haar measure in the lattice path integral
\begin{equation}
  \label{eq:pathint}
  Z = \int [\scrD U] [\scrD X] \ \pf(\cM) \ e^{-S_B},
\end{equation}
and do not preserve any exact supersymmetries at non-zero lattice spacing $a > 0$.
The RHMC algorithm can encounter instabilities when the coupling $g$ is too strong, in a way that depends on both the lattice action as well as the lattice spacing set by $N_{\tau}$.
For the simple lattice action above, and considering our coarsest lattice spacing with $N_{\tau} = 8$, we encounter instabilities for $g \gsim 0.025$, significantly smaller than the values reached by Refs.~\cite{Asano:2018nol, Bergner:2021goh}.
This may motivate switching to a more complicated improved action in future work.

\begin{figure}[tbp]
  \includegraphics[width=\linewidth]{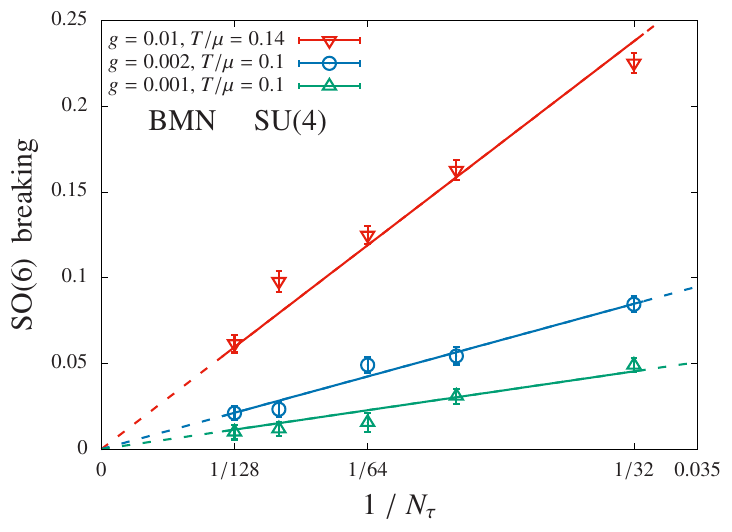}
  \caption{\label{fig:SO6_breaking}Confirming that the SO(6) symmetry breaking observed with our lattice formulation is a discretization artifact that vanishes in the $N_{\tau} \to \infty$ continuum limit.  These tests consider gauge group SU(4) to enable computations with large $N_{\tau} \leq 128$ for three different $g = 0.001$, $0.002$ and $0.01$.}
\end{figure}

Lattice discretization introduces $a$-dependent artifacts in numerical calculations, which also depend on the lattice action and typically increase with $g$.
In particular, we have observed that for our simple lattice action, discretization artifacts break the expected SO(6) symmetry~\cite{Schaich:2020ubh, Schaich:2022duk}.
Specifically, the six gauge-invariant $\Tr{X_M^2}$ split into a set of two with larger values and a set of four with smaller values --- still significantly larger than the three $\Tr{X_I^2}$.
To quantify this effect, we consider the ratio
\begin{equation}
  \label{eq:splitting}
  R_{\text{SO(6)}} \equiv \frac{\vev{\Tr{X_{(2)}^2}} - \vev{\Tr{X_{(4)}^2}}}{\vev{\Tr{X_{(6)}^2}}},
\end{equation}
where $\vev{\Tr{X_{(2)}^2}}$, $\vev{\Tr{X_{(4)}^2}}$, and $\vev{\Tr{X_{(6)}^2}}$ average over the two larger traces, the four smaller traces and all six of them, respectively.
In \fig{fig:SO6_breaking}, we plot this ratio for three different couplings $g = 0.001$, $0.002$, and $0.01$, considering a small SU(4) gauge group to access large lattice sizes up to $N_{\tau} = 128$ close to the continuum limit.
We find that the SO(6) breaking vanishes linearly in the $N_\tau \to \infty$ continuum limit, confirming that it is merely a discretization artifact, which may be reduced or removed by employing an improved lattice action in future work.

\begin{figure}[tbp]
  \includegraphics[width=\linewidth]{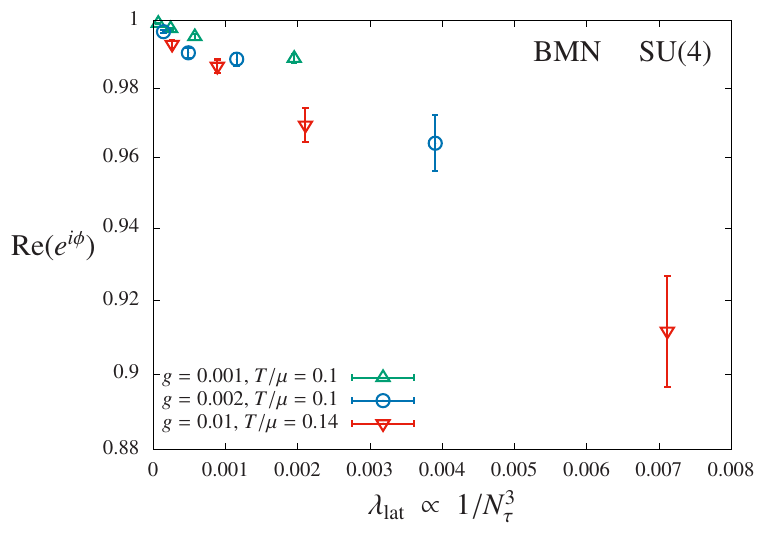}
  \caption{\label{fig:pfaffian_N4}The real part of $\vev{e^{i\phi}}_{\text{pq}}$ plotted vs.\ $\lalat \propto 1 / N_\tau^3$ for gauge group SU(4) confirms that the Pfaffian becomes real and positive in the $N_\tau \to \infty$ continuum limit where $\lalat \to 0$.}
\end{figure}
\begin{figure}[tbp]
  \includegraphics[width=\linewidth]{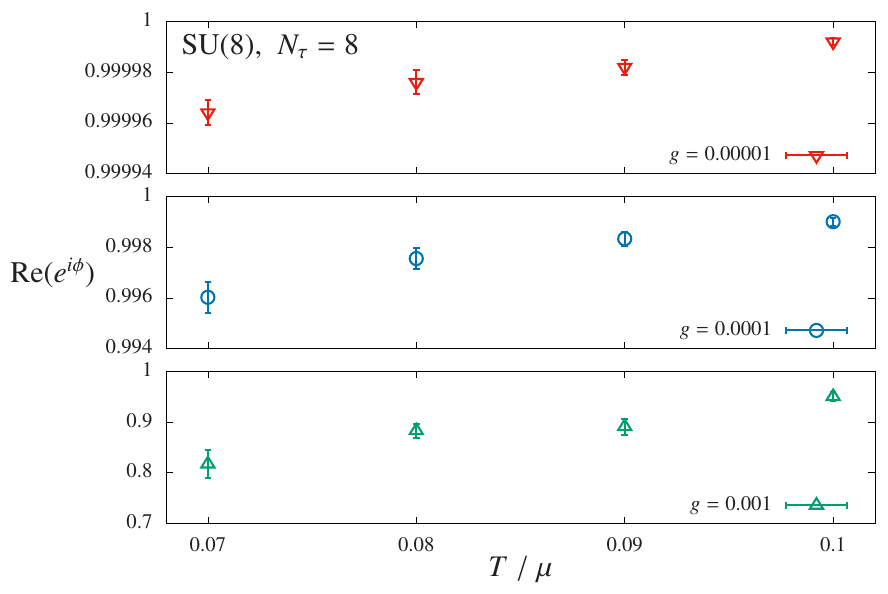}
  \caption{\label{fig:pfaffian_N8}The real part of $\vev{e^{i\phi}}_{\text{pq}}$ plotted vs.\ $T / \mu$ for gauge group SU(8) with $N_\tau = 8$ and three values of $g = 10^{-5}$, $10^{-4}$ and $0.001$ (from top to bottom).  The Pfaffian phase fluctuations increase as $g$ increases and as $T / \mu$ decreases, which is likely related to instabilities in calculations with $g \gsim 0.025$ for this small $N_{\tau} = 8$.}
\end{figure}

One other complication is that the RHMC algorithm treats the factor of $\pf(\cM) e^{-S_B}$ in the path integral \eq{eq:pathint} as a real, positive Boltzmann weight.
However, our lattice discretization of the BMN model allows the Pfaffian to be complex, $\pf(\cM) = |\pf(\cM)| e^{i\phi}$.
We proceed by `quenching' the phase $e^{i\phi} \to 1$~\cite{Schaich:2014pda}, which in principle requires reweighting in order to recover the true expectation values $\vev{\cO}$ from phase-quenched (`$_{\text{pq}}$') calculations.
That is,
\begin{align}
  \vev{\cO} = \frac{\int [\scrD U] [\scrD X] \ \cO e^{-S_B}\ \pf(\cM)}{\int [\scrD U] [\scrD X] \ e^{-S_B} \ \pf(\cM)} = \frac{\vev{\cO e^{i\phi}}_{\text{pq}}}{\vev{e^{i\phi}}_{\text{pq}}}, & \\
  \mbox{where}~\vev{\cO}_{\text{pq}} = \frac{\int [\scrD U] [\scrD X] \ \cO e^{-S_B}\ |\pf(\cM)|}{\int [\scrD U] [\scrD X] \ e^{-S_B} \ |\pf(\cM)|}. &
\end{align}
Reweighting requires computationally expensive measurements of the Pfaffian phase $\vev{e^{i\phi}}_{\text{pq}}$ and fails if this expectation value is consistent with zero (a `sign problem').

In Figs.~\ref{fig:pfaffian_N4} and \ref{fig:pfaffian_N8}, we present some checks of the phase of the Pfaffian $\vev{e^{i\phi}}_{\text{pq}}$, which show that our phase-quenched numerical results are not significantly affected by phase reweighting.
In \fig{fig:pfaffian_N4} we again consider gauge group SU(4), plotting the real part of $\vev{e^{i\phi}}_{\text{pq}}$ against $\lalat = g / [(T / \mu) N_{\tau}]^3$.
The right-most point for each data set corresponds to $N_{\tau} = 8$, while the largest systems we consider here have $N_{\tau} = 24$.
For $N_{\tau} = 24$, the Pfaffian phase measurement for a single field configuration is roughly 300 times more computationally expensive than generating a single molecular dynamics time unit (MDTU) with the RHMC algorithm.
The plot confirms that the Pfaffian becomes real and positive in the $N_{\tau} \to \infty$ continuum limit, with no sign problem for sufficiently large lattices.

Finally, \fig{fig:pfaffian_N8} shows the same quantity for some SU(8) systems similar to (but smaller than) those we analyze to determine the phase diagram in the next section.
Here $N_{\tau} = 8$ and the computational cost for each Pfaffian phase measurement increases to roughly 3000 times the cost of generating an MDTU with the RHMC algorithm.
We consider three values of $g = 10^{-5}$, $10^{-4}$, and $0.001$ (from top to bottom) and plot the results against values of $T / \mu$ ranging from the confined phase to the deconfined phase.
We can clearly see that the Pfaffian phase fluctuations increase as the coupling increases and as the temperature decreases.
Because the right-most green point for $g = 0.001$ matches the corresponding point in \fig{fig:pfaffian_N4} apart from the different gauge group, we can also check that the fluctuations increase from $1 - \vev{e^{i\phi}}_{\text{pq}} = 0.0113(14)$ for $N = 4$ to $0.0501(84)$ for $N = 8$.
These Pfaffian phase fluctuations are likely related to the instabilities we observe for $N_{\tau} = 8$ and $g \gsim 0.025$, motivating us to impose a minimum $N_{\tau} \geq 16$ for strong couplings $g \geq 0.001$ in our main calculations, to which we now turn.

\section{Phase Diagram Results}
\label{sec:results}

We organize our lattice investigations of the deconfinement transition in terms of the dimensionless parameters $g$ and $T / \mu$ discussed above, \eq{eq:params}.
We consider four fixed values of the coupling that span three orders of magnitude: $g = 10^{-5}$, $10^{-4}$, 0.001, and 0.01.
In each case, we determine the transition temperature by scanning in the ratio $T / \mu$, generating $\cO(10)$ ensembles that range from the confined phase to the deconfined phase.
As discussed at the end of \secref{sec:BMN_Matrix_Model}, this differs from the approaches taken by prior studies~\cite{Catterall:2010gf, Asano:2018nol, Bergner:2021goh}, most of which also employ different lattice actions.
To explore the thermodynamic limit where the number of colors $N \to \infty$, for each $g$ we repeat this procedure for up to three values of $N = 8$, 12, and 16.
To explore the continuum limit where the number of lattice sites $N_{\tau} \to \infty$, for each $(g, N)$ we consider two lattice sizes, either $N_{\tau} = 8$ and $16$ for the two weaker couplings $g \leq 10^{-4}$ or $N_{\tau} = 16$ and $24$ for the two stronger couplings $g \geq 0.001$.
\tab{tab:results} in \appref{app:data} summarizes these details.
In total, these calculations involve 261 ensembles.
Full information about them and the 35 additional SU(4) and SU(8) ensembles considered in the previous section is available in our open data release~\cite{data}.

To study the deconfinement transition of the BMN model, we focus on the Polyakov loop
\begin{equation}
  \label{eq:poly}
  \vev{|PL|} = \frac{1}{N} \vev{\left| \Tr{\prod_{n = 0}^{N_{\tau} - 1} U_{\tau}(n)}\right|} \equiv \frac{1}{N} \vev{|\Tr{\Pbb}|}.
\end{equation}
That is, by ``Polyakov loop'' we refer specifically to the magnitude of the trace of the holonomy along the temporal circle, with that holonomy itself corresponding to the $N \X N$ matrix $\Pbb$.
In the large-$N$ limit, $\vev{|PL|}$ is an order parameter for the spontaneous breaking of the $Z_N$ center symmetry, which vanishes in the confined phase while remaining non-zero in the deconfined phase.

\begin{figure*}[tbp]
  \includegraphics[width=0.45\linewidth]{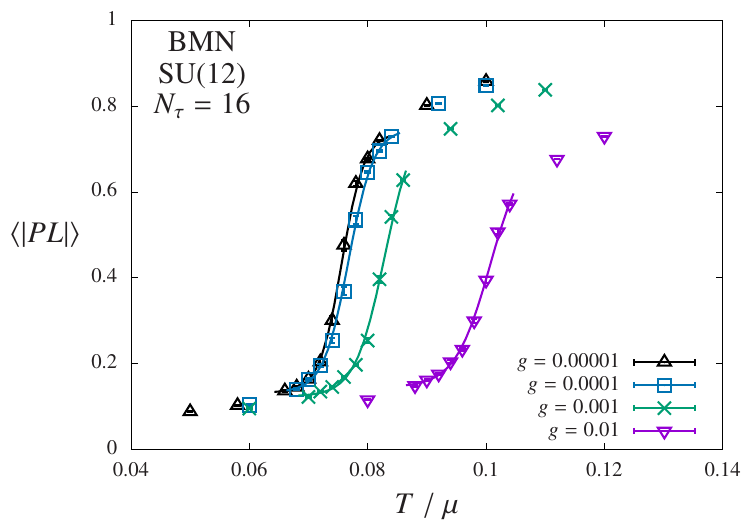}\hfill
  \includegraphics[width=0.45\linewidth]{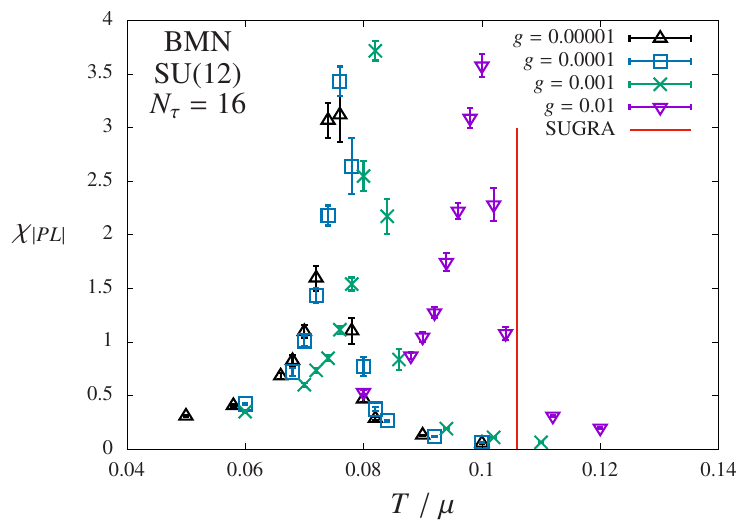}
  \caption{\label{fig:poly_all_g} Results for the Polyakov loop (left) and its susceptibility (right) plotted against $T / \mu$ and computed with a fixed number of sites $N_\tau = 16$ for gauge group SU(12). The four data sets correspond to four values of the coupling $g$ spanning three orders of magnitude. The transition moves to larger \critinline as $g$ increases. The lines on the left plot are fits to the sigmoid ansatz in \eq{eq:sigmoid}.  The vertical red line in the right plot indicates the classical supergravity prediction in the large-$N$ continuum $g \to \infty$ limit, \eq{eq:strong}.}
\end{figure*}
\begin{figure*}[tbp]
  \includegraphics[width=0.45\linewidth]{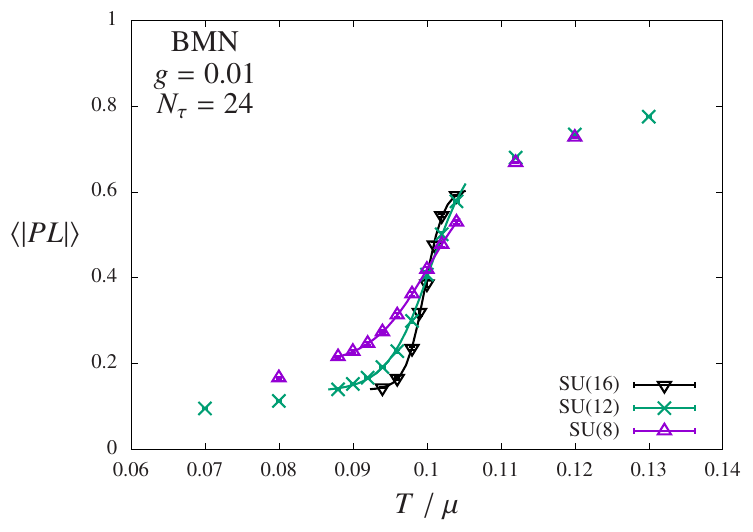}\hfill
  \includegraphics[width=0.45\linewidth]{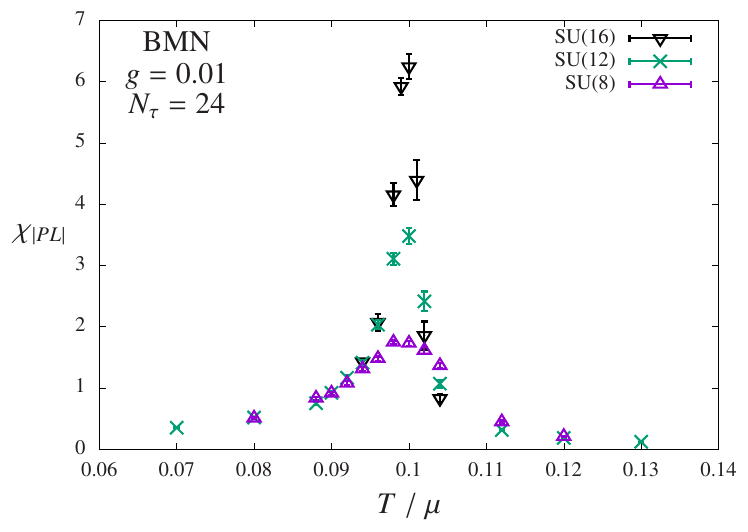}
  \caption{\label{fig:poly_all_N}The Polyakov loop (left) and its susceptibility (right) as in \fig{fig:poly_all_g}, now fixing $N_{\tau} = 24$ and $g = 0.01$ with three data sets corresponding to different numbers of colors $N = 8$, $12$, and $16$. The susceptibility peak at the critical temperature becomes higher as $N$ increases towards the thermodynamic limit.}
\end{figure*}

This behavior is illustrated by the left-hand plots in Figs.~\ref{fig:poly_all_g} and \ref{fig:poly_all_N}, which show the Polyakov loop increasing as $T / \mu$ increases from the confined to the deconfined phase.
Figure~\ref{fig:poly_all_g} compares all four values of the coupling $g$ for fixed $N_{\tau} = 16$ and gauge group SU(12), while \fig{fig:poly_all_N} compares the three gauge groups for fixed $N_{\tau} = 24$ and $g = 0.01$.
(There is very little dependence on $N_{\tau}$ with $g$ and $N$ fixed, which can be seen from \tab{tab:results} in \appref{app:data}.)
The right-hand plots in these figures present the corresponding Polyakov loop susceptibility
\begin{equation}
  \chi_{_{|PL|}} \equiv \vev{|PL|^2} - \vev{|PL|}^2,
\end{equation}
which exhibits a peak at the critical temperature \critinline of the deconfinement transition.
By eye, \fig{fig:poly_all_g} clearly shows the expected behavior, with \critinline moving from around the weak-coupling limit \eq{eq:weak} towards the strong-coupling supergravity prediction \eq{eq:strong} (marked by the vertical red line) as the coupling increases.
Recall that the latter value corresponds to the large-$N$, $g \to \infty$ limit of the continuum theory and is not a strict bound on our finite-$N$ calculations.

We can obtain simple estimates for \critinline from the locations of the susceptibility peaks.
These estimates are collected in \tab{tab:results}, but are limited by the discrete values of $T / \mu$ we have analyzed.
In principle, this can be improved by using multi-ensemble reweighting to interpolate between these values~\cite{Kuramashi:2020meg, Ferrenberg:1988yz}.
Here, we instead take a simpler approach of interpolating our results for the Polyakov loop itself.
Specifically, we fit $\vev{|PL|}$ to the four-parameter sigmoid ansatz~\cite{Schaich:2020ubh, Schaich:2022duk}
\begin{equation}
  \label{eq:sigmoid}
  \Sigma = A - \frac{B}{1 + \exp\left[C (T / \mu - D)\right]}.
\end{equation}
All four parameters are necessary since our finite-$N$ Polyakov loop results are non-zero for all temperatures.
We also need to omit the highest- and lowest-temperature points from the fits in order to keep the $\chidof$ under control.
These fits correspond to the solid lines in the left-hand plots of Figs.~\ref{fig:poly_all_g} and \ref{fig:poly_all_N}, which extend only as far as the subset of points that are included.

The transition temperature \critinline is given by the fit parameter $D$ that corresponds to the inflection point where the sigmoid's slope is maximized:
\begin{align*}
  \frac{d^2 \Sigma}{d(T / \mu)^2} \propto 1 - \frac{2\exp\left[C (T / \mu - D)\right]}{1 + \exp\left[C (T / \mu - D)\right]} & = 0 \\
  \implies C (T / \mu - D) & = 0.
\end{align*}
As shown by \tab{tab:results} in \appref{app:data}, our fits produce values for \critinline with small uncertainties, which are fully consistent with the susceptibility peaks.
However, \tab{tab:results} also shows that the fits typically involve uncomfortably large $\chidof$, suggesting that these small uncertainties are likely underestimated.
We therefore proceed by using the central values from the sigmoid fits but the more conservative uncertainties from the susceptibility peaks.
This approach should account for systematic effects (e.g., from the choices of ansatz and fit range) in addition to purely statistical uncertainties.

\begin{figure}[tbp]
  \centering
  \includegraphics[width=\linewidth]{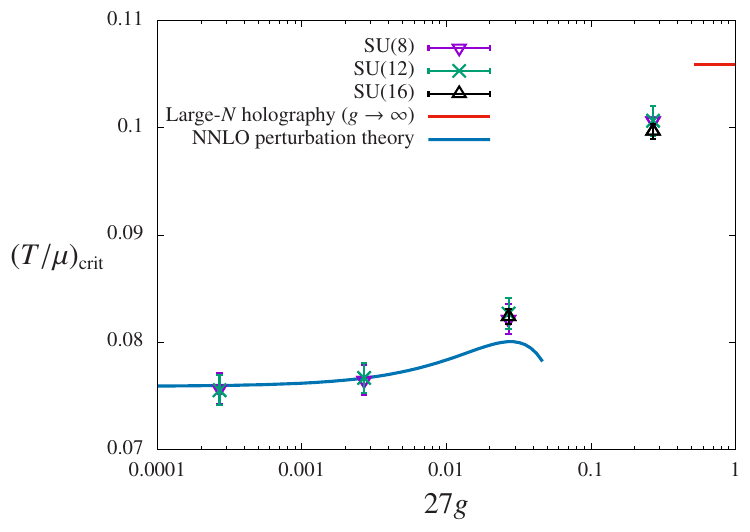}
  \caption{\label{fig:phase_diag}The critical temperature of the transition, $\critinline$, vs.\ the coupling $g$ on semi-log axes, from $N_{\tau} \to \infty$ extrapolations described in the text. The horizontal red line is the $g \to \infty$ supergravity prediction from \refcite{Costa:2014wya} while the blue curve is the NNLO perturbative prediction (with expansion parameter $27g$) in \eq{eq:NNLO}.}
\end{figure}

The results we obtain in this way are used to produce the \critinline vs.\ $g$ phase diagram shown in \fig{fig:phase_diag}.
While we have only two lattice sizes $N_{\tau}$ for each $\{N, g\}$, from \tab{tab:results} in \appref{app:data} we can see that all these pairs of \critinline results agree within uncertainties.
We therefore average each pair by fitting to a constant, which leaves a single degree of freedom and produces very small $0.005 \lsim \chidof \lsim 0.3$.
Figure~\ref{fig:phase_diag} shows these averages for each $\{N, g\}$, making it clear that there is also no visible $N$-dependence in our results.

At weak coupling $g \lsim 10^{-4}$ we find critical temperatures in excellent agreement with NNLO perturbation theory in the $N \to \infty$ planar limit, \eq{eq:NNLO}.
Because we use the perturbative expansion parameter $27g$ on the horizontal axis, these points appear at $27g \lsim 0.003$.
Perturbation theory breaks down around $27g \approx 0.05$ due to the relatively large coefficient $C_{\text{NNLO}} \approx 71$ in \eq{eq:NNLO}.
Our non-perturbative numerical results monotonically increase as $g$ increases, approaching the large-$N$, strong-coupling supergravity prediction from \eq{eq:strong}, which is marked by a horizontal red line in \fig{fig:phase_diag}.
In fact, it is remarkable how close we get to this $g \to \infty$ limit given our modest $g \leq 0.01$.
If we were able to reach stronger couplings, we might observe our finite-$N$ lattice calculations exceeding the large-$N$ supergravity prediction, as in \refcite{Bergner:2021goh}, but this will require future work to implement an improved lattice action.
In parallel, our results also provide motivation for dual-supergravity calculations to attempt to determine the functional dependence of \critinline on $g$, to compare with the numerical results we have obtained.

\begin{figure*}[tbp]
  \includegraphics[width=0.45\linewidth]{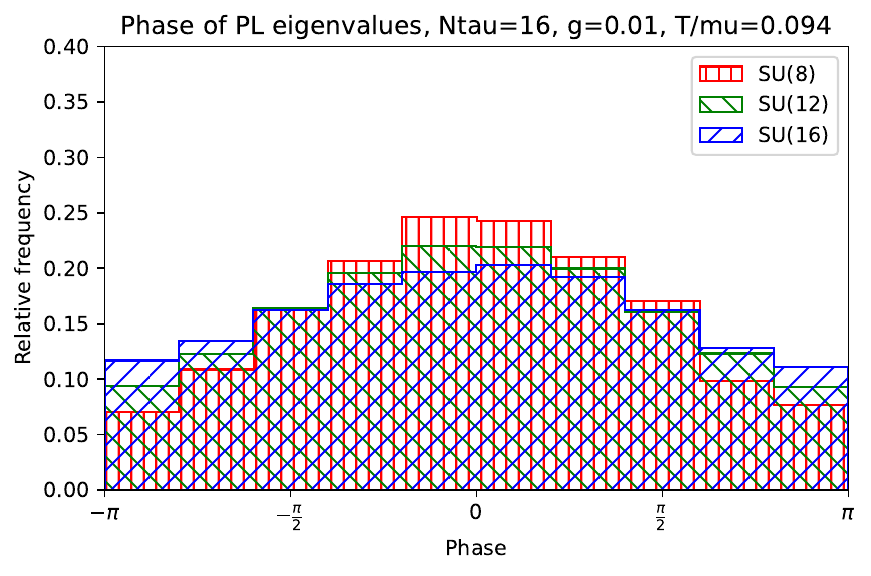}\hfill \includegraphics[width=0.45\linewidth]{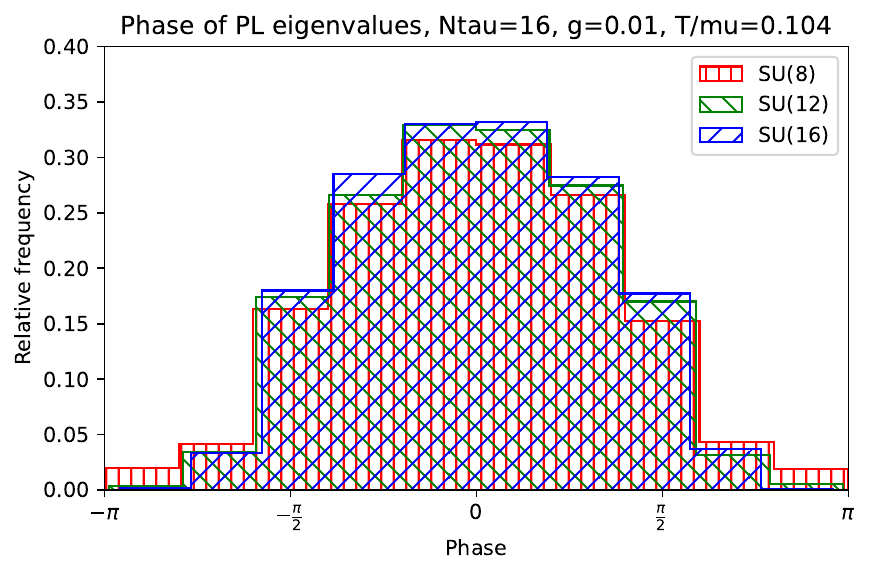}
  \caption{\label{fig:PLeig}Angular distributions of Polyakov loop eigenvalue phases for $N_\tau = 16$ and $g = 0.01$, comparing SU($N$) gauge groups with $N = 8$, $12$ and $16$. \textbf{Left:} For $T / \mu = 0.094$ below the transition, the distributions become broader as $N$ increases, consistent with the uniform distribution expected for the confined phase in the large-$N$ limit. \textbf{Right:} For $T / \mu = 0.104$ above the transition, the distributions become more localized as $N$ increases, as expected for the deconfined phase.}
\end{figure*}

Although we have now presented our main results for the BMN model phase diagram in \fig{fig:phase_diag}, there are two more analyses we can carry out based on the Polyakov loop.
First, we confirm that we are, in fact, dealing with the expected transition between confined and deconfined phases (as opposed to the `fuzzy-sphere' vacua corresponding to \eq{eq:fuzzy}) by considering the eigenvalues of the $N \X N$ matrix $\Pbb$, \eq{eq:poly}.
In the high-temperature deconfined phase, these $N$ eigenvalues for a given field configuration should all be aligned around one of the $Z_N$ vacua, with a localized distribution of complex phases.
In the low-temperature confined phase, we should have instead a uniform distribution of eigenvalue phases around the unit circle~\cite{Maldacena:1998im, Witten:1998zw, Aharony:2004ig}.
While it is possible to estimate \critinline (in the large-$N$ limit) by determining the temperature at which the eigenvalue phase distribution first spreads out to cover $[-\pi, \pi)$ with no gap remaining~\cite{Aharony:2003sx, Aharony:2004ig}, here we simply check systems on either side of the transition.
Figure~\ref{fig:PLeig} shows a representative check, for $N_{\tau} = 16$ and $g = 0.01$, comparing SU($N$) gauge groups with $N = 8$, $12$ and $16$.
For $T / \mu = 0.094$ below the transition, we see that the distributions become broader as $N$ increases, consistent with the uniform distribution expected for the confined phase in the large-$N$ limit.
For $T / \mu = 0.104$ above the transition, the distributions become more localized as $N$ increases, as expected for the deconfined phase.

\begin{figure}[tbp]
  \includegraphics[width=\linewidth]{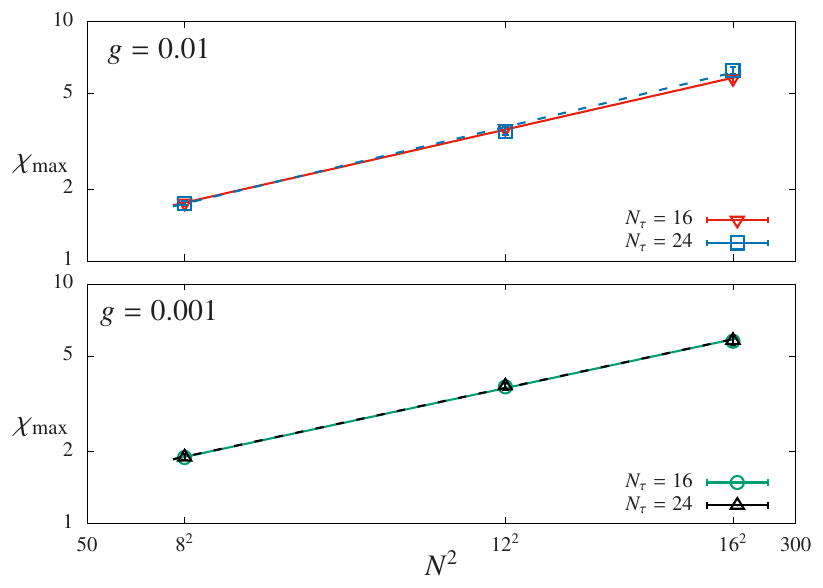}
  \caption{\label{fig:suscept_scaling}The scaling of the susceptibility peak \chimax with the number of degrees of freedom $\propto N^2$ on log--log axes.  The solid and dashed lines are power-law fits with fixed $N_{\tau} = 16$ and $24$, respectively.}
\end{figure}

Second and finally, we study the scaling of the Polyakov loop susceptibility peak \chimax with the number of degrees of freedom $\propto N^2$, which can distinguish between first-order and continuous transitions.
Already in \fig{fig:poly_all_N}, we can see that the peak becomes higher as $N$ increases towards the $N \to \infty$ thermodynamic limit.
In \fig{fig:suscept_scaling}, we plot \chimax against $N^2$ on log--log axes for couplings $g = 0.01$ (top) and $g = 0.001$ (bottom), and fit the data to a power law,
\begin{equation}
  \chimax = C N^{2b},
\end{equation}
where $C$ and $b$ are fit parameters.
For the strongest coupling $g = 0.01$, the critical exponent is $b = 0.866(29)$ for $N_{\tau} = 16$, increasing to $0.909(26)$ for $N_{\tau} = 24$ (with only statistical uncertainties).
These results are consistent with the first-order value $b = 1$~\cite{Imry:1980zz, Fisher:1982xt, Binder:1984llk, Challa:1986sk, Fukugita:1989yb} in the continuum limit $N_{\tau} \to \infty$.
The $g = 0.001$ results are also close to one, although with no dependence on $N_{\tau}$ at all: $b = 0.816(32)$ and $0.815(34)$ for $N_{\tau} = 16$ and $24$, respectively.
These fits leave a single degree of freedom and feature reasonable $0.2 \lsim \chidof \lsim 1.7$.

Using perturbative calculations to compute an effective action, \refcite{Hadizadeh:2004bf} suggests that the BMN phase transition remains first order in the weakly coupled regime.
For $g \leq 10^{-4}$, we can report only rougher results due to the smaller $N_{\tau} = 8$ and $16$ we consider (cf.\ \tab{tab:results}), each with only two gauge groups SU(8) and SU(12).
Still, these \critinline allow us to estimate the critical exponent (neglecting uncertainties), and we observe that it now becomes smaller as $N_{\tau}$ increases towards the continuum limit.
Specifically, for $g = 10^{-4}$ we find that $b$ decreases from $0.85$ to $0.61$ as $N_{\tau}$ increases from $8$ to $16$.
Similarly, for $g = 10^{-5}$, $b \simeq 0.71$ for $N_{\tau} = 8$ decreases to $0.59$ for $N_{\tau} = 16$.
These results suggest that although we have a first-order transition for strong couplings $g \gtrsim 0.001$, the transition appears to be continuous for weaker couplings $g \lesssim 10^{-4}$.
Future work is needed to confirm this and search for a possible critical endpoint $g_{\star}$ to the line of strong-coupling first-order transitions.

\section{Conclusions}
\label{sec:conc}

We have presented results from our numerical Monte Carlo investigations of the BMN matrix model on the lattice, featuring our determination of the critical temperature of the model's deconfinement phase transition for dimensionless couplings $g$ that span three orders of magnitude from the perturbative regime towards the domain of the dual supergravity, \fig{fig:phase_diag}.
Considering multiple SU($N$) gauge groups and lattice sizes $N_{\tau}$ for each $g$, we reproduce \critinline from perturbation theory for $g \lsim 10^{-4}$ while approaching the large-$N$ strong-coupling limit from gauge/gravity duality~\cite{Costa:2014wya} for $g \gsim 0.001$.
We have also analyzed the order of the transition, finding that the first-order transition for strong couplings appears to become continuous for weaker $g \lsim 10^{-4}$.

While our results appear broadly consistent with previous lattice studies of the BMN model's phase diagram~\cite{Catterall:2010gf, Asano:2018nol, Bergner:2021goh}, our approach differs by both our choice of lattice discretization as well as our strategy for analyzing the system.
In particular, discretization-dependent instabilities in RHMC calculations restrict us to couplings $g \lsim 0.025$ significantly smaller than those reached by Refs.~\cite{Asano:2018nol, Bergner:2021goh}.
We have discussed related discretization artifacts, including the phase of the fermion operator Pfaffian, which we observe to fluctuate significantly at strong coupling and large lattice spacing, \fig{fig:pfaffian_N8}.
Despite this, a sign problem can be avoided by working with sufficiently small lattice spacings, which make the Pfaffian real and positive, \fig{fig:pfaffian_N4}.

Given the relatively small $g \leq 0.01$ we consider, it is remarkable how close our results in \fig{fig:phase_diag} get to the $g \to \infty$ limit from \refcite{Costa:2014wya}.
As discussed at the end of \secref{sec:BMN_Matrix_Model}, it is possible that finite-$N$ corrections may allow \critinline to exceed this strong-coupling limit~\cite{Bergner:2021goh}.
This motivates further calculations with stronger couplings as well as larger values of $N$ in order to carry out better-controlled extrapolations to the $N^2 \to \infty$ thermodynamic limit.
From our investigations of discretization artifacts and the Pfaffian in \secref{sec:latform}, we can appreciate that this will require either (or both) analyzing larger lattice sizes $N_{\tau}$ or switching to an improved lattice action.
Larger $N_{\tau}$ will also enable better-controlled continuum extrapolations than were possible in this work.
In addition, considering larger SU($N$) gauge groups is also important at weaker couplings to improve the scaling analyses that suggest the deconfinement transition becomes continuous rather than first order.
If this observation is confirmed, it will be interesting to search for the critical endpoint $g_{\star}$ to the line of first-order transitions at stronger couplings.

In general, the BMN model remains less explored than the $\mu \to 0$ BFSS limit, and there are further investigations we may pursue using numerical lattice Monte Carlo calculations.
These include computations of the internal energy of the system as a function of temperature~\cite{Pateloudis:2022ijr}, as well as analyses of the additional `fuzzy-sphere' vacua corresponding to \eq{eq:fuzzy}~\cite{Asano:2018nol} or the effects of `ungauging' the system~\cite{Pateloudis:2022oos}.

\section*{Acknowledgements}

We thank Simon Catterall and Toby Wiseman for helpful conversations, Denjoe O'Connor for useful discussions at the ECT* workshop ``Quantum Gravity meets Lattice QFT'', and Yuhma Asano for sharing numerical data from \refcite{Asano:2018nol}.
RGJ was supported by the U.S.\ Department of Energy, Office of Science, Office of Nuclear Physics under contract {DE-AC05-06OR23177}.
AJ was supported in part by the Start-up Research Grant from the University of the Witwatersrand, South Africa.
DS was supported by UK Research and Innovation Future Leader Fellowship {MR/S015418/1} \& {MR/X015157/1} and STFC grants {ST/T000988/1} \& {ST/X000699/1}.
Additional support came from the International Centre for Theoretical Sciences Bangalore (ICTS) during visits to participate in the two editions of the program ``Nonperturbative and Numerical Approaches to Quantum Gravity, String Theory and Holography'' ({ICTS/Prog-NUMSTRINGS2018/01}, {ICTS/numstrings-2022/8}); we thank ICTS for its hospitality.
Numerical calculations were carried out at the University of Liverpool, on USQCD facilities at Fermilab funded by the U.S.\ Department of Energy, and at the San Diego Supercomputer Center and Pittsburgh Supercomputing Center through XSEDE supported by National Science Foundation grant number {ACI-1548562}. \\[8 pt]

\noindent \textbf{Data Availability Statement:} All data used in this work are available through the open data release \refcite{data}, which also provides the bulk of the computational workflow needed to reproduce, check, and extend our analyses.

\begin{appendix}

\section{\label{app:gamma}Gamma matrices}
In order to construct the lattice action for the BMN model, we need to choose a basis to represent the $16 \X 16$ matrices $\ga_i$, $\ga_{\tau}$ and $\ga_{123}$. 
In terms of the real $2\times 2$ matrices
\begin{align*}
  X & \equiv \si_1   = \begin{pmatrix}0 & 1 \\ 1 & 0\end{pmatrix} \\
  Y & \equiv -i\si_2 = \begin{pmatrix}0 & -1 \\ 1 & 0\end{pmatrix} \\
  Z & \equiv \si_3   = \begin{pmatrix}1 & 0 \\ 0 & -1\end{pmatrix} \\
  1 & \equiv \Ibb_2  = \begin{pmatrix}1 & 0 \\ 0 & 1\end{pmatrix},
\end{align*}
the representation we employ can be written as the following Kronecker products:
\begin{align*}
  \ga_1      & = Y \otimes Y \otimes Y \otimes Y \\
  \ga_2      & = Y \otimes X \otimes Y \otimes 1 \\
  \ga_3      & = Y \otimes Z \otimes Y \otimes 1 \\
  \ga_{123}  & = Y \otimes X \otimes Y \otimes Y \\
  \ga_4      & = Y \otimes Y \otimes 1 \otimes X \\
  \ga_5      & = Y \otimes Y \otimes 1 \otimes Z \\
  \ga_6      & = Y \otimes 1 \otimes X \otimes Y \\
  \ga_7      & = Y \otimes 1 \otimes Z \otimes Y \\
  \ga_8      & = -Z \otimes 1 \otimes 1 \otimes 1 \\
  \ga_9      & = -i1 \otimes 1 \otimes 1 \otimes 1 \\
  \ga_{\tau} & = X \otimes 1 \otimes 1 \otimes 1.
\end{align*}
These expressions match the discussion around \eq{eq:gamma10} and can be checked against the explicit implementation in \refcite{susy_code}.

\section{\label{app:data}Numerical results}

\begin{table}[tbp]
  \centering
  \renewcommand\arraystretch{1.3}   
  \addtolength{\tabcolsep}{1 pt}    
  \begin{tabular}{c|cc|c|c|l|c}
    $g$       & $N$ & $N_{\tau}$ & $\sharp$ ens. & $\max(\chi)$ & Sigmoid fit  & \chidof \\\hline\hline 
    0.01      &  8  & 16         & 19            & 0.100(2)     & 0.10098(19)  &  1.7    \\             
    0.01      &  8  & 24         & 12            & 0.098(2)     & 0.10029(64)  &  0.01   \\\hline       
    0.01      & 12  & 16         & 12            & 0.100(2)     & 0.10073(31)  &  4.6    \\             
    0.01      & 12  & 24         & 14            & 0.100(2)     & 0.10054(27)  &  3.0    \\\hline       
    0.01      & 16  & 16         &  8            & 0.099(1)     & 0.099598(58) &  4.9    \\             
    0.01      & 16  & 24         &  8            & 0.100(1)     & 0.099714(48) &  2.0    \\\hline       
    \hline             
    0.001     &  8  & 16         & 21            & 0.082(2)     & 0.082917(81) & 13      \\             
    0.001     &  8  & 24         & 12            & 0.082(2)     & 0.08135(52)  &  5.5    \\\hline       
    0.001     & 12  & 16         & 13            & 0.082(2)     & 0.08285(19)  & 15      \\             
    0.001     & 12  & 24         & 14            & 0.082(2)     & 0.082448(44) & 54      \\\hline       
    0.001     & 16  & 16         & 10            & 0.082(1)     & 0.082445(31) & 19      \\             
    0.001     & 16  & 24         & 10            & 0.082(1)     & 0.082251(35) & 19      \\\hline       
    \hline             
    $10^{-4}$ &  8  &  8         & 16            & 0.076(2)     & 0.07602(18)  & 16      \\             
    $10^{-4}$ &  8  & 16         & 14            & 0.076(2)     & 0.07684(19)  &  8.9    \\\hline       
    $10^{-4}$ & 12  &  8         & 13            & 0.076(2)     & 0.076504(63) &  7.9    \\             
    $10^{-4}$ & 12  & 16         & 12            & 0.076(2)     & 0.076767(62) &  6.8    \\\hline       
    \hline             
    $10^{-5}$ &  8  &  8         & 14            & 0.074(2)     & 0.075435(63) & 28      \\             
    $10^{-5}$ &  8  & 16         & 13            & 0.076(2)     & 0.075856(98) &  7.8    \\\hline       
    $10^{-5}$ & 12  &  8         & 13            & 0.074(2)     & 0.075352(83) &  3.6    \\             
    $10^{-5}$ & 12  & 16         & 13            & 0.076(2)     & 0.075631(71) &  5.1    \\\hline       
  \end{tabular}
  \caption{\label{tab:results}Summary of the 261 SU(8), SU(12) and SU(16) ensembles used in our lattice analyses of the BMN model phase diagram, with two determinations of the critical transition temperature \critinline for each $\{g, N, N_{\tau}\}$.  The fifth column reports \critinline corresponding to the maximum Polyakov loop susceptibility, while the sixth column considers instead the fit to the sigmoid ansatz \eq{eq:sigmoid} discussed in \secref{sec:results}.  Both estimates agree within uncertainties, with the generally large \chidof of the sigmoid fits suggesting the corresponding uncertainties on \critinline are underestimated.  Full information, including 35 additional SU(4) and SU(8) ensembles, is available in \refcite{data}.}
\end{table}

\tab{tab:results} summarizes the 261 lattice ensembles used in our numerical analyses of the BMN model's phase diagram, including the results for the critical temperatures \critinline obtained from both susceptibility peaks as well as sigmoid fits to the Polyakov loop itself, \eq{eq:sigmoid}.
The final column reports the \chidof of the corresponding sigmoid fit.
For each value of the dimensionless coupling $g$ spanning three orders of magnitude, we consider multiple SU($N$) gauge groups up to $N = 16$ and two lattice sizes up to $N_{\tau} = 24$.
In every case, we use the RHMC algorithm~\cite{Clark:2006fx} implemented by our publicly available parallel software for lattice supersymmetry~\cite{susy_code, Schaich:2014pda} to generate $\cO(10)$ ensembles that range from the confined phase to the deconfined phase.
For each ensemble, we typically generate 5000 MDTU, using unit-length trajectories in the RHMC algorithm, and impose a thermalization cut after human inspection of automated time-series plots.
We generated and analyzed 35 additional ensembles with gauge groups SU(4) and SU(8) to carry out the checks of discretization artifacts and the Pfaffian phase discussed in \secref{sec:latform}.

For each ensemble, simple observables including the Polyakov loop, the Myers term \eq{eq:Myers_lat} and the gauge-invariant `scalar squares' $\Tr{X_i^2}$ are measured after every RHMC trajectory of length 1~MDTU.
These data confirm that the `fuzzy-sphere' vacua corresponding to \eq{eq:fuzzy} do not play a role in our current numerical investigations.
More involved computations of the extremal eigenvalues of the squared fermion operator $\cMdag \cM$ are done using configurations saved to disk every 10~MDTU.
These eigenvalue computations are performed using a Davidson-type method provided by the PReconditioned Iterative Multi-Method Eigensolver (PRIMME) library~\cite{Stathopoulos:2010}.
We require that the extremal eigenvalues remain within the spectral range where the rational approximation used in the RHMC algorithm is reliable.

After generating configurations and carrying out extremal eigenvalue measurements, we use the `autocorr' module in \texttt{emcee}~\cite{Foreman:2013mc} to estimate auto-correlation times $\tau$ for three relevant quantities.
These are the magnitude of the Polyakov loop, the lowest eigenvalue of $\cMdag \cM$, and a representative scalar square, $\Tr{X_9^2}$.
We divide our thermalized measurements into blocks for jackknife analyses, confirming that our default block size of 100~MDTU is always larger than all three auto-correlation times.
Our approach ensures that at least 31 statistically independent blocks are available for each ensemble, more than enough for robust analyses.
For the 27 SU(4) and SU(8) ensembles we use to carry out more expensive measurements of the fermion operator Pfaffian, we compute $\pf(\cM)$ only once per jackknife block, in other words using every tenth saved configuration after the thermalization cut.

For all 296 ensembles, all of this information and more is presented in our open data release \refcite{data}, which also provides the bulk of the computational workflow needed to reproduce, check, and extend our analyses.

\end{appendix}

\raggedright
\bibliography{BMN_phase}

\begin{thebibliography}{73}%
\makeatletter
\providecommand \@ifxundefined [1]{%
 \@ifx{#1\undefined}
}%
\providecommand \@ifnum [1]{%
 \ifnum #1\expandafter \@firstoftwo
 \else \expandafter \@secondoftwo
 \fi
}%
\providecommand \@ifx [1]{%
 \ifx #1\expandafter \@firstoftwo
 \else \expandafter \@secondoftwo
 \fi
}%
\providecommand \natexlab [1]{#1}%
\providecommand \enquote  [1]{``#1''}%
\providecommand \bibnamefont  [1]{#1}%
\providecommand \bibfnamefont [1]{#1}%
\providecommand \citenamefont [1]{#1}%
\providecommand \href@noop [0]{\@secondoftwo}%
\providecommand \href [0]{\begingroup \@sanitize@url \@href}%
\providecommand \@href[1]{\@@startlink{#1}\@@href}%
\providecommand \@@href[1]{\endgroup#1\@@endlink}%
\providecommand \@sanitize@url [0]{\catcode `\\12\catcode `\$12\catcode
  `\&12\catcode `\#12\catcode `\^12\catcode `\_12\catcode `\%12\relax}%
\providecommand \@@startlink[1]{}%
\providecommand \@@endlink[0]{}%
\providecommand \url  [0]{\begingroup\@sanitize@url \@url }%
\providecommand \@url [1]{\endgroup\@href {#1}{\urlprefix }}%
\providecommand \urlprefix  [0]{URL }%
\providecommand \Eprint [0]{\href }%
\providecommand \doibase [0]{http://dx.doi.org/}%
\providecommand \selectlanguage [0]{\@gobble}%
\providecommand \bibinfo  [0]{\@secondoftwo}%
\providecommand \bibfield  [0]{\@secondoftwo}%
\providecommand \translation [1]{[#1]}%
\providecommand \BibitemOpen [0]{}%
\providecommand \bibitemStop [0]{}%
\providecommand \bibitemNoStop [0]{.\EOS\space}%
\providecommand \EOS [0]{\spacefactor3000\relax}%
\providecommand \BibitemShut  [1]{\csname bibitem#1\endcsname}%
\let\auto@bib@innerbib\@empty
\bibitem [{\citenamefont {Itzhaki}\ \emph {et~al.}(1998)\citenamefont
  {Itzhaki}, \citenamefont {Maldacena}, \citenamefont {Sonnenschein},\ and\
  \citenamefont {Yankielowicz}}]{Itzhaki:1998dd}%
  \BibitemOpen
  \bibfield  {author} {\bibinfo {author} {\bibfnamefont {N.}~\bibnamefont
  {Itzhaki}}, \bibinfo {author} {\bibfnamefont {J.~M.}\ \bibnamefont
  {Maldacena}}, \bibinfo {author} {\bibfnamefont {J.}~\bibnamefont
  {Sonnenschein}}, \ and\ \bibinfo {author} {\bibfnamefont {S.}~\bibnamefont
  {Yankielowicz}},\ }\bibfield  {title} {\enquote {\bibinfo {title}
  {{Supergravity and the large-$N$ limit of theories with sixteen
  supercharges}},}\ }\href {\doibase 10.1103/PhysRevD.58.046004} {\bibfield
  {journal} {\bibinfo  {journal} {Phys. Rev. D}\ }\textbf {\bibinfo {volume}
  {58}},\ \bibinfo {pages} {046004} (\bibinfo {year} {1998})},\ \Eprint
  {http://arxiv.org/abs/hep-th/9802042} {hep-th/9802042} \BibitemShut {NoStop}%
\bibitem [{\citenamefont {Maldacena}(1998{\natexlab{a}})}]{Maldacena:1997re}%
  \BibitemOpen
  \bibfield  {author} {\bibinfo {author} {\bibfnamefont {J.~M.}\ \bibnamefont
  {Maldacena}},\ }\bibfield  {title} {\enquote {\bibinfo {title} {{The
  Large-$N$ limit of superconformal field theories and supergravity}},}\ }\href
  {\doibase 10.4310/ATMP.1998.v2.n2.a1} {\bibfield  {journal} {\bibinfo
  {journal} {Adv. Theor. Math. Phys.}\ }\textbf {\bibinfo {volume} {2}},\
  \bibinfo {pages} {231--252} (\bibinfo {year} {1998}{\natexlab{a}})},\ \Eprint
  {http://arxiv.org/abs/hep-th/9711200} {hep-th/9711200} \BibitemShut {NoStop}%
\bibitem [{\citenamefont {Catterall}\ and\ \citenamefont
  {Wiseman}(2007)}]{Catterall:2007fp}%
  \BibitemOpen
  \bibfield  {author} {\bibinfo {author} {\bibfnamefont {S.}~\bibnamefont
  {Catterall}}\ and\ \bibinfo {author} {\bibfnamefont {T.}~\bibnamefont
  {Wiseman}},\ }\bibfield  {title} {\enquote {\bibinfo {title} {{Towards
  lattice simulation of the gauge theory duals to black holes and hot
  strings}},}\ }\href {\doibase 10.1088/1126-6708/2007/12/104} {\bibfield
  {journal} {\bibinfo  {journal} {JHEP}\ }\textbf {\bibinfo {volume} {0712}},\
  \bibinfo {pages} {104} (\bibinfo {year} {2007})},\ \Eprint
  {http://arxiv.org/abs/0706.3518} {arXiv:0706.3518} \BibitemShut {NoStop}%
\bibitem [{\citenamefont {Anagnostopoulos}\ \emph {et~al.}(2008)\citenamefont
  {Anagnostopoulos}, \citenamefont {Hanada}, \citenamefont {Nishimura},\ and\
  \citenamefont {Takeuchi}}]{Anagnostopoulos:2007fw}%
  \BibitemOpen
  \bibfield  {author} {\bibinfo {author} {\bibfnamefont {K.~N.}\ \bibnamefont
  {Anagnostopoulos}}, \bibinfo {author} {\bibfnamefont {M.}~\bibnamefont
  {Hanada}}, \bibinfo {author} {\bibfnamefont {J.}~\bibnamefont {Nishimura}}, \
  and\ \bibinfo {author} {\bibfnamefont {S.}~\bibnamefont {Takeuchi}},\
  }\bibfield  {title} {\enquote {\bibinfo {title} {{Monte Carlo studies of
  supersymmetric matrix quantum mechanics with sixteen supercharges at finite
  temperature}},}\ }\href {\doibase 10.1103/PhysRevLett.100.021601} {\bibfield
  {journal} {\bibinfo  {journal} {Phys. Rev. Lett.}\ }\textbf {\bibinfo
  {volume} {100}},\ \bibinfo {pages} {021601} (\bibinfo {year} {2008})},\
  \Eprint {http://arxiv.org/abs/0707.4454} {arXiv:0707.4454} \BibitemShut
  {NoStop}%
\bibitem [{\citenamefont {Catterall}\ and\ \citenamefont
  {Wiseman}(2008)}]{Catterall:2008yz}%
  \BibitemOpen
  \bibfield  {author} {\bibinfo {author} {\bibfnamefont {S.}~\bibnamefont
  {Catterall}}\ and\ \bibinfo {author} {\bibfnamefont {T.}~\bibnamefont
  {Wiseman}},\ }\bibfield  {title} {\enquote {\bibinfo {title} {{Black hole
  thermodynamics from simulations of lattice Yang--Mills theory}},}\ }\href
  {\doibase 10.1103/PhysRevD.78.041502} {\bibfield  {journal} {\bibinfo
  {journal} {Phys. Rev. D}\ }\textbf {\bibinfo {volume} {78}},\ \bibinfo
  {pages} {041502} (\bibinfo {year} {2008})},\ \Eprint
  {http://arxiv.org/abs/0803.4273} {arXiv:0803.4273} \BibitemShut {NoStop}%
\bibitem [{\citenamefont {Catterall}(2009)}]{Catterall:2008dv}%
  \BibitemOpen
  \bibfield  {author} {\bibinfo {author} {\bibfnamefont {S.}~\bibnamefont
  {Catterall}},\ }\bibfield  {title} {\enquote {\bibinfo {title} {{First
  results from simulations of supersymmetric lattices}},}\ }\href {\doibase
  10.1088/1126-6708/2009/01/040} {\bibfield  {journal} {\bibinfo  {journal}
  {JHEP}\ }\textbf {\bibinfo {volume} {0901}},\ \bibinfo {pages} {040}
  (\bibinfo {year} {2009})},\ \Eprint {http://arxiv.org/abs/0811.1203}
  {arXiv:0811.1203} \BibitemShut {NoStop}%
\bibitem [{\citenamefont {Hanada}\ \emph
  {et~al.}(2009{\natexlab{a}})\citenamefont {Hanada}, \citenamefont {Miwa},
  \citenamefont {Nishimura},\ and\ \citenamefont {Takeuchi}}]{Hanada:2008gy}%
  \BibitemOpen
  \bibfield  {author} {\bibinfo {author} {\bibfnamefont {M.}~\bibnamefont
  {Hanada}}, \bibinfo {author} {\bibfnamefont {A.}~\bibnamefont {Miwa}},
  \bibinfo {author} {\bibfnamefont {J.}~\bibnamefont {Nishimura}}, \ and\
  \bibinfo {author} {\bibfnamefont {S.}~\bibnamefont {Takeuchi}},\ }\bibfield
  {title} {\enquote {\bibinfo {title} {{Schwarzschild radius from Monte Carlo
  calculation of the Wilson loop in supersymmetric matrix quantum
  mechanics}},}\ }\href {\doibase 10.1103/PhysRevLett.102.181602} {\bibfield
  {journal} {\bibinfo  {journal} {Phys. Rev. Lett.}\ }\textbf {\bibinfo
  {volume} {102}},\ \bibinfo {pages} {181602} (\bibinfo {year}
  {2009}{\natexlab{a}})},\ \Eprint {http://arxiv.org/abs/0811.2081}
  {arXiv:0811.2081} \BibitemShut {NoStop}%
\bibitem [{\citenamefont {Hanada}\ \emph
  {et~al.}(2009{\natexlab{b}})\citenamefont {Hanada}, \citenamefont
  {Hyakutake}, \citenamefont {Nishimura},\ and\ \citenamefont
  {Takeuchi}}]{Hanada:2008ez}%
  \BibitemOpen
  \bibfield  {author} {\bibinfo {author} {\bibfnamefont {M.}~\bibnamefont
  {Hanada}}, \bibinfo {author} {\bibfnamefont {Y.}~\bibnamefont {Hyakutake}},
  \bibinfo {author} {\bibfnamefont {J.}~\bibnamefont {Nishimura}}, \ and\
  \bibinfo {author} {\bibfnamefont {S.}~\bibnamefont {Takeuchi}},\ }\bibfield
  {title} {\enquote {\bibinfo {title} {{Higher Derivative Corrections to Black
  Hole Thermodynamics from Supersymmetric Matrix Quantum Mechanics}},}\ }\href
  {\doibase 10.1103/PhysRevLett.102.191602} {\bibfield  {journal} {\bibinfo
  {journal} {Phys. Rev. Lett.}\ }\textbf {\bibinfo {volume} {102}},\ \bibinfo
  {pages} {191602} (\bibinfo {year} {2009}{\natexlab{b}})},\ \Eprint
  {http://arxiv.org/abs/0811.3102} {arXiv:0811.3102} \BibitemShut {NoStop}%
\bibitem [{\citenamefont {Catterall}\ and\ \citenamefont
  {Wiseman}(2010)}]{Catterall:2009xn}%
  \BibitemOpen
  \bibfield  {author} {\bibinfo {author} {\bibfnamefont {S.}~\bibnamefont
  {Catterall}}\ and\ \bibinfo {author} {\bibfnamefont {T.}~\bibnamefont
  {Wiseman}},\ }\bibfield  {title} {\enquote {\bibinfo {title} {{Extracting
  black hole physics from the lattice}},}\ }\href {\doibase
  10.1007/JHEP04(2010)077} {\bibfield  {journal} {\bibinfo  {journal} {JHEP}\
  }\textbf {\bibinfo {volume} {1004}},\ \bibinfo {pages} {077} (\bibinfo {year}
  {2010})},\ \Eprint {http://arxiv.org/abs/0909.4947} {arXiv:0909.4947}
  \BibitemShut {NoStop}%
\bibitem [{\citenamefont {Nishimura}(2009)}]{Nishimura:2009xm}%
  \BibitemOpen
  \bibfield  {author} {\bibinfo {author} {\bibfnamefont {J.}~\bibnamefont
  {Nishimura}},\ }\bibfield  {title} {\enquote {\bibinfo {title} {{Non-lattice
  simulation of supersymmetric gauge theories as a probe to quantum black holes
  and strings}},}\ }\href {\doibase 10.22323/1.091.0016} {\bibfield  {journal}
  {\bibinfo  {journal} {Proc. Sci.}\ }\textbf {\bibinfo {volume} {LAT2009}},\
  \bibinfo {pages} {016} (\bibinfo {year} {2009})},\ \Eprint
  {http://arxiv.org/abs/0912.0327} {arXiv:0912.0327} \BibitemShut {NoStop}%
\bibitem [{\citenamefont {Catterall}\ and\ \citenamefont {van
  Anders}(2010)}]{Catterall:2010gf}%
  \BibitemOpen
  \bibfield  {author} {\bibinfo {author} {\bibfnamefont {S.}~\bibnamefont
  {Catterall}}\ and\ \bibinfo {author} {\bibfnamefont {G.}~\bibnamefont {van
  Anders}},\ }\bibfield  {title} {\enquote {\bibinfo {title} {{First Results
  from Lattice Simulation of the PWMM}},}\ }\href {\doibase
  10.1007/JHEP09(2010)088} {\bibfield  {journal} {\bibinfo  {journal} {JHEP}\
  }\textbf {\bibinfo {volume} {1009}},\ \bibinfo {pages} {088} (\bibinfo {year}
  {2010})},\ \Eprint {http://arxiv.org/abs/1003.4952} {arXiv:1003.4952}
  \BibitemShut {NoStop}%
\bibitem [{\citenamefont {Catterall}\ \emph {et~al.}(2010)\citenamefont
  {Catterall}, \citenamefont {Joseph},\ and\ \citenamefont
  {Wiseman}}]{Catterall:2010fx}%
  \BibitemOpen
  \bibfield  {author} {\bibinfo {author} {\bibfnamefont {S.}~\bibnamefont
  {Catterall}}, \bibinfo {author} {\bibfnamefont {A.}~\bibnamefont {Joseph}}, \
  and\ \bibinfo {author} {\bibfnamefont {T.}~\bibnamefont {Wiseman}},\
  }\bibfield  {title} {\enquote {\bibinfo {title} {{Thermal phases of D1-branes
  on a circle from lattice super-Yang--Mills}},}\ }\href {\doibase
  10.1007/JHEP12(2010)022} {\bibfield  {journal} {\bibinfo  {journal} {JHEP}\
  }\textbf {\bibinfo {volume} {1012}},\ \bibinfo {pages} {022} (\bibinfo {year}
  {2010})},\ \Eprint {http://arxiv.org/abs/1008.4964} {arXiv:1008.4964}
  \BibitemShut {NoStop}%
\bibitem [{\citenamefont {Catterall}\ \emph {et~al.}(2012)\citenamefont
  {Catterall}, \citenamefont {Galvez}, \citenamefont {Joseph},\ and\
  \citenamefont {Mehta}}]{Catterall:2011aa}%
  \BibitemOpen
  \bibfield  {author} {\bibinfo {author} {\bibfnamefont {S.}~\bibnamefont
  {Catterall}}, \bibinfo {author} {\bibfnamefont {R.}~\bibnamefont {Galvez}},
  \bibinfo {author} {\bibfnamefont {A.}~\bibnamefont {Joseph}}, \ and\ \bibinfo
  {author} {\bibfnamefont {D.}~\bibnamefont {Mehta}},\ }\bibfield  {title}
  {\enquote {\bibinfo {title} {{On the sign problem in 2D lattice
  super-Yang--Mills}},}\ }\href {\doibase 10.1007/JHEP01(2012)108} {\bibfield
  {journal} {\bibinfo  {journal} {JHEP}\ }\textbf {\bibinfo {volume} {1201}},\
  \bibinfo {pages} {108} (\bibinfo {year} {2012})},\ \Eprint
  {http://arxiv.org/abs/1112.3588} {arXiv:1112.3588} \BibitemShut {NoStop}%
\bibitem [{\citenamefont {Kadoh}\ and\ \citenamefont
  {Kamata}(2012)}]{Kadoh:2012bg}%
  \BibitemOpen
  \bibfield  {author} {\bibinfo {author} {\bibfnamefont {D.}~\bibnamefont
  {Kadoh}}\ and\ \bibinfo {author} {\bibfnamefont {S.}~\bibnamefont {Kamata}},\
  }\bibfield  {title} {\enquote {\bibinfo {title} {{One-dimensional
  supersymmetric Yang--Mills theory with 16 supercharges}},}\ }\href {\doibase
  10.22323/1.164.0064} {\bibfield  {journal} {\bibinfo  {journal} {Proc. Sci.}\
  }\textbf {\bibinfo {volume} {LATTICE2012}},\ \bibinfo {pages} {064} (\bibinfo
  {year} {2012})},\ \Eprint {http://arxiv.org/abs/1212.4919} {arXiv:1212.4919}
  \BibitemShut {NoStop}%
\bibitem [{\citenamefont {Hanada}\ \emph {et~al.}(2014)\citenamefont {Hanada},
  \citenamefont {Hyakutake}, \citenamefont {Ishiki},\ and\ \citenamefont
  {Nishimura}}]{Hanada:2013rga}%
  \BibitemOpen
  \bibfield  {author} {\bibinfo {author} {\bibfnamefont {M.}~\bibnamefont
  {Hanada}}, \bibinfo {author} {\bibfnamefont {Y.}~\bibnamefont {Hyakutake}},
  \bibinfo {author} {\bibfnamefont {G.}~\bibnamefont {Ishiki}}, \ and\ \bibinfo
  {author} {\bibfnamefont {J.}~\bibnamefont {Nishimura}},\ }\bibfield  {title}
  {\enquote {\bibinfo {title} {{Holographic description of quantum black hole
  on a computer}},}\ }\href {\doibase 10.1126/science.1250122} {\bibfield
  {journal} {\bibinfo  {journal} {Science}\ }\textbf {\bibinfo {volume}
  {344}},\ \bibinfo {pages} {882--885} (\bibinfo {year} {2014})},\ \Eprint
  {http://arxiv.org/abs/1311.5607} {arXiv:1311.5607} \BibitemShut {NoStop}%
\bibitem [{\citenamefont {Gigu{\`e}re}\ and\ \citenamefont
  {Kadoh}(2015)}]{Giguere:2015cga}%
  \BibitemOpen
  \bibfield  {author} {\bibinfo {author} {\bibfnamefont {E.}~\bibnamefont
  {Gigu{\`e}re}}\ and\ \bibinfo {author} {\bibfnamefont {D.}~\bibnamefont
  {Kadoh}},\ }\bibfield  {title} {\enquote {\bibinfo {title} {{Restoration of
  supersymmetry in two-dimensional SYM with sixteen supercharges on the
  lattice}},}\ }\href {\doibase 10.1007/JHEP05(2015)082} {\bibfield  {journal}
  {\bibinfo  {journal} {JHEP}\ }\textbf {\bibinfo {volume} {1505}},\ \bibinfo
  {pages} {082} (\bibinfo {year} {2015})},\ \Eprint
  {http://arxiv.org/abs/1503.04416} {arXiv:1503.04416} \BibitemShut {NoStop}%
\bibitem [{\citenamefont {Kadoh}\ and\ \citenamefont
  {Kamata}(2015)}]{Kadoh:2015mka}%
  \BibitemOpen
  \bibfield  {author} {\bibinfo {author} {\bibfnamefont {D.}~\bibnamefont
  {Kadoh}}\ and\ \bibinfo {author} {\bibfnamefont {S.}~\bibnamefont {Kamata}},\
  }\bibfield  {title} {\enquote {\bibinfo {title} {{Gauge/gravity duality and
  lattice simulations of one-dimensional SYM with sixteen supercharges}},}\
  }\href@noop {} {\  (\bibinfo {year} {2015})},\ \Eprint
  {http://arxiv.org/abs/1503.08499} {arXiv:1503.08499} \BibitemShut {NoStop}%
\bibitem [{\citenamefont {Filev}\ and\ \citenamefont
  {O'Connor}(2016)}]{Filev:2015hia}%
  \BibitemOpen
  \bibfield  {author} {\bibinfo {author} {\bibfnamefont {V.~G.}\ \bibnamefont
  {Filev}}\ and\ \bibinfo {author} {\bibfnamefont {D.}~\bibnamefont
  {O'Connor}},\ }\bibfield  {title} {\enquote {\bibinfo {title} {{The BFSS
  model on the lattice}},}\ }\href {\doibase 10.1007/JHEP05(2016)167}
  {\bibfield  {journal} {\bibinfo  {journal} {JHEP}\ }\textbf {\bibinfo
  {volume} {1605}},\ \bibinfo {pages} {167} (\bibinfo {year} {2016})},\ \Eprint
  {http://arxiv.org/abs/1506.01366} {arXiv:1506.01366} \BibitemShut {NoStop}%
\bibitem [{\citenamefont {Hanada}\ \emph {et~al.}(2016)\citenamefont {Hanada},
  \citenamefont {Hyakutake}, \citenamefont {Ishiki},\ and\ \citenamefont
  {Nishimura}}]{Hanada:2016zxj}%
  \BibitemOpen
  \bibfield  {author} {\bibinfo {author} {\bibfnamefont {M.}~\bibnamefont
  {Hanada}}, \bibinfo {author} {\bibfnamefont {Y.}~\bibnamefont {Hyakutake}},
  \bibinfo {author} {\bibfnamefont {G.}~\bibnamefont {Ishiki}}, \ and\ \bibinfo
  {author} {\bibfnamefont {J.}~\bibnamefont {Nishimura}},\ }\bibfield  {title}
  {\enquote {\bibinfo {title} {{Numerical tests of the gauge/gravity duality
  conjecture for D0-branes at finite temperature and finite $N$}},}\ }\href
  {\doibase 10.1103/PhysRevD.94.086010} {\bibfield  {journal} {\bibinfo
  {journal} {Phys. Rev. D}\ }\textbf {\bibinfo {volume} {94}},\ \bibinfo
  {pages} {086010} (\bibinfo {year} {2016})},\ \Eprint
  {http://arxiv.org/abs/1603.00538} {arXiv:1603.00538} \BibitemShut {NoStop}%
\bibitem [{\citenamefont {Berkowitz}\ \emph
  {et~al.}(2016{\natexlab{a}})\citenamefont {Berkowitz}, \citenamefont
  {Rinaldi}, \citenamefont {Hanada}, \citenamefont {Ishiki}, \citenamefont
  {Shimasaki},\ and\ \citenamefont {Vranas}}]{Berkowitz:2016tyy}%
  \BibitemOpen
  \bibfield  {author} {\bibinfo {author} {\bibfnamefont {E.}~\bibnamefont
  {Berkowitz}}, \bibinfo {author} {\bibfnamefont {E.}~\bibnamefont {Rinaldi}},
  \bibinfo {author} {\bibfnamefont {M.}~\bibnamefont {Hanada}}, \bibinfo
  {author} {\bibfnamefont {G.}~\bibnamefont {Ishiki}}, \bibinfo {author}
  {\bibfnamefont {S.}~\bibnamefont {Shimasaki}}, \ and\ \bibinfo {author}
  {\bibfnamefont {P.}~\bibnamefont {Vranas}},\ }\bibfield  {title} {\enquote
  {\bibinfo {title} {{Supergravity from D0-brane Quantum Mechanics}},}\
  }\href@noop {} {\  (\bibinfo {year} {2016}{\natexlab{a}})},\ \Eprint
  {http://arxiv.org/abs/1606.04948} {arXiv:1606.04948} \BibitemShut {NoStop}%
\bibitem [{\citenamefont {Berkowitz}\ \emph
  {et~al.}(2016{\natexlab{b}})\citenamefont {Berkowitz}, \citenamefont
  {Rinaldi}, \citenamefont {Hanada}, \citenamefont {Ishiki}, \citenamefont
  {Shimasaki},\ and\ \citenamefont {Vranas}}]{Berkowitz:2016jlq}%
  \BibitemOpen
  \bibfield  {author} {\bibinfo {author} {\bibfnamefont {E.}~\bibnamefont
  {Berkowitz}}, \bibinfo {author} {\bibfnamefont {E.}~\bibnamefont {Rinaldi}},
  \bibinfo {author} {\bibfnamefont {M.}~\bibnamefont {Hanada}}, \bibinfo
  {author} {\bibfnamefont {G.}~\bibnamefont {Ishiki}}, \bibinfo {author}
  {\bibfnamefont {S.}~\bibnamefont {Shimasaki}}, \ and\ \bibinfo {author}
  {\bibfnamefont {P.}~\bibnamefont {Vranas}},\ }\bibfield  {title} {\enquote
  {\bibinfo {title} {{Precision lattice test of the gauge/gravity duality at
  large $N$}},}\ }\href {\doibase 10.1103/PhysRevD.94.094501} {\bibfield
  {journal} {\bibinfo  {journal} {Phys. Rev. D}\ }\textbf {\bibinfo {volume}
  {94}},\ \bibinfo {pages} {094501} (\bibinfo {year} {2016}{\natexlab{b}})},\
  \Eprint {http://arxiv.org/abs/1606.04951} {arXiv:1606.04951} \BibitemShut
  {NoStop}%
\bibitem [{\citenamefont {Kadoh}(2017)}]{Kadoh:2017mcj}%
  \BibitemOpen
  \bibfield  {author} {\bibinfo {author} {\bibfnamefont {D.}~\bibnamefont
  {Kadoh}},\ }\bibfield  {title} {\enquote {\bibinfo {title} {{Precision test
  of the gauge/gravity duality in two-dimensional $\mathcal N = (8, 8)$
  SYM}},}\ }\href {\doibase 10.22323/1.256.0033} {\bibfield  {journal}
  {\bibinfo  {journal} {Proc. Sci.}\ }\textbf {\bibinfo {volume}
  {LATTICE2016}},\ \bibinfo {pages} {033} (\bibinfo {year} {2017})},\ \Eprint
  {http://arxiv.org/abs/1702.01615} {arXiv:1702.01615} \BibitemShut {NoStop}%
\bibitem [{\citenamefont {Rinaldi}\ \emph {et~al.}(2018)\citenamefont
  {Rinaldi}, \citenamefont {Berkowitz}, \citenamefont {Hanada}, \citenamefont
  {Maltz},\ and\ \citenamefont {Vranas}}]{Rinaldi:2017mjl}%
  \BibitemOpen
  \bibfield  {author} {\bibinfo {author} {\bibfnamefont {E.}~\bibnamefont
  {Rinaldi}}, \bibinfo {author} {\bibfnamefont {E.}~\bibnamefont {Berkowitz}},
  \bibinfo {author} {\bibfnamefont {M.}~\bibnamefont {Hanada}}, \bibinfo
  {author} {\bibfnamefont {J.}~\bibnamefont {Maltz}}, \ and\ \bibinfo {author}
  {\bibfnamefont {P.}~\bibnamefont {Vranas}},\ }\bibfield  {title} {\enquote
  {\bibinfo {title} {{Toward Holographic Reconstruction of Bulk Geometry from
  Lattice Simulations}},}\ }\href {\doibase 10.1007/JHEP02(2018)042} {\bibfield
   {journal} {\bibinfo  {journal} {JHEP}\ }\textbf {\bibinfo {volume} {1802}},\
  \bibinfo {pages} {042} (\bibinfo {year} {2018})},\ \Eprint
  {http://arxiv.org/abs/1709.01932} {arXiv:1709.01932} \BibitemShut {NoStop}%
\bibitem [{\citenamefont {Catterall}\ \emph {et~al.}(2018)\citenamefont
  {Catterall}, \citenamefont {Jha}, \citenamefont {Schaich},\ and\
  \citenamefont {Wiseman}}]{Catterall:2017lub}%
  \BibitemOpen
  \bibfield  {author} {\bibinfo {author} {\bibfnamefont {S.}~\bibnamefont
  {Catterall}}, \bibinfo {author} {\bibfnamefont {R.~G.}\ \bibnamefont {Jha}},
  \bibinfo {author} {\bibfnamefont {D.}~\bibnamefont {Schaich}}, \ and\
  \bibinfo {author} {\bibfnamefont {T.}~\bibnamefont {Wiseman}},\ }\bibfield
  {title} {\enquote {\bibinfo {title} {{Testing holography using lattice
  super-Yang--Mills theory on a 2-torus}},}\ }\href {\doibase
  10.1103/PhysRevD.97.086020} {\bibfield  {journal} {\bibinfo  {journal} {Phys.
  Rev. D}\ }\textbf {\bibinfo {volume} {97}},\ \bibinfo {pages} {086020}
  (\bibinfo {year} {2018})},\ \Eprint {http://arxiv.org/abs/1709.07025}
  {arXiv:1709.07025} \BibitemShut {NoStop}%
\bibitem [{\citenamefont {Jha}\ \emph {et~al.}(2018)\citenamefont {Jha},
  \citenamefont {Catterall}, \citenamefont {Schaich},\ and\ \citenamefont
  {Wiseman}}]{Jha:2017zad}%
  \BibitemOpen
  \bibfield  {author} {\bibinfo {author} {\bibfnamefont {R.~G.}\ \bibnamefont
  {Jha}}, \bibinfo {author} {\bibfnamefont {S.}~\bibnamefont {Catterall}},
  \bibinfo {author} {\bibfnamefont {D.}~\bibnamefont {Schaich}}, \ and\
  \bibinfo {author} {\bibfnamefont {T.}~\bibnamefont {Wiseman}},\ }\bibfield
  {title} {\enquote {\bibinfo {title} {{Testing the holographic principle using
  lattice simulations}},}\ }\href {\doibase 10.1051/epjconf/201817508004}
  {\bibfield  {journal} {\bibinfo  {journal} {EPJ Web Conf.}\ }\textbf
  {\bibinfo {volume} {175}},\ \bibinfo {pages} {08004} (\bibinfo {year}
  {2018})},\ \Eprint {http://arxiv.org/abs/1710.06398} {arXiv:1710.06398}
  \BibitemShut {NoStop}%
\bibitem [{\citenamefont {Berkowitz}\ \emph {et~al.}(2018)\citenamefont
  {Berkowitz}, \citenamefont {Hanada}, \citenamefont {Rinaldi},\ and\
  \citenamefont {Vranas}}]{Berkowitz:2018qhn}%
  \BibitemOpen
  \bibfield  {author} {\bibinfo {author} {\bibfnamefont {E.}~\bibnamefont
  {Berkowitz}}, \bibinfo {author} {\bibfnamefont {M.}~\bibnamefont {Hanada}},
  \bibinfo {author} {\bibfnamefont {E.}~\bibnamefont {Rinaldi}}, \ and\
  \bibinfo {author} {\bibfnamefont {P.}~\bibnamefont {Vranas}},\ }\bibfield
  {title} {\enquote {\bibinfo {title} {{Gauged And Ungauged: A Nonperturbative
  Test}},}\ }\href {\doibase 10.1007/JHEP06(2018)124} {\bibfield  {journal}
  {\bibinfo  {journal} {JHEP}\ }\textbf {\bibinfo {volume} {1806}},\ \bibinfo
  {pages} {124} (\bibinfo {year} {2018})},\ \Eprint
  {http://arxiv.org/abs/1802.02985} {arXiv:1802.02985} \BibitemShut {NoStop}%
\bibitem [{\citenamefont {Asano}\ \emph {et~al.}(2018)\citenamefont {Asano},
  \citenamefont {Filev}, \citenamefont {Kov{\'a}{\v c}ik},\ and\ \citenamefont
  {O'Connor}}]{Asano:2018nol}%
  \BibitemOpen
  \bibfield  {author} {\bibinfo {author} {\bibfnamefont {Y.}~\bibnamefont
  {Asano}}, \bibinfo {author} {\bibfnamefont {V.~G.}\ \bibnamefont {Filev}},
  \bibinfo {author} {\bibfnamefont {S.}~\bibnamefont {Kov{\'a}{\v c}ik}}, \
  and\ \bibinfo {author} {\bibfnamefont {D.}~\bibnamefont {O'Connor}},\
  }\bibfield  {title} {\enquote {\bibinfo {title} {{The non-perturbative phase
  diagram of the BMN matrix model}},}\ }\href {\doibase
  10.1007/JHEP07(2018)152} {\bibfield  {journal} {\bibinfo  {journal} {JHEP}\
  }\textbf {\bibinfo {volume} {1807}},\ \bibinfo {pages} {152} (\bibinfo {year}
  {2018})},\ \Eprint {http://arxiv.org/abs/1805.05314} {arXiv:1805.05314}
  \BibitemShut {NoStop}%
\bibitem [{\citenamefont {Schaich}\ \emph {et~al.}(2020)\citenamefont
  {Schaich}, \citenamefont {Jha},\ and\ \citenamefont
  {Joseph}}]{Schaich:2020ubh}%
  \BibitemOpen
  \bibfield  {author} {\bibinfo {author} {\bibfnamefont {D.}~\bibnamefont
  {Schaich}}, \bibinfo {author} {\bibfnamefont {R.~G.}\ \bibnamefont {Jha}}, \
  and\ \bibinfo {author} {\bibfnamefont {A.}~\bibnamefont {Joseph}},\
  }\bibfield  {title} {\enquote {\bibinfo {title} {{Thermal phase structure of
  a supersymmetric matrix model}},}\ }\href {\doibase 10.22323/1.363.0069}
  {\bibfield  {journal} {\bibinfo  {journal} {Proc. Sci.}\ }\textbf {\bibinfo
  {volume} {LATTICE2019}},\ \bibinfo {pages} {069} (\bibinfo {year} {2020})},\
  \Eprint {http://arxiv.org/abs/2003.01298} {arXiv:2003.01298} \BibitemShut
  {NoStop}%
\bibitem [{\citenamefont {Bergner}\ \emph
  {et~al.}(2022{\natexlab{a}})\citenamefont {Bergner}, \citenamefont
  {Bodendorfer}, \citenamefont {Hanada}, \citenamefont {Pateloudis},
  \citenamefont {Rinaldi}, \citenamefont {Sch{\"a}fer}, \citenamefont
  {Vranas},\ and\ \citenamefont {Watanabe}}]{Bergner:2021goh}%
  \BibitemOpen
  \bibfield  {author} {\bibinfo {author} {\bibfnamefont {G.}~\bibnamefont
  {Bergner}}, \bibinfo {author} {\bibfnamefont {N.}~\bibnamefont
  {Bodendorfer}}, \bibinfo {author} {\bibfnamefont {M.}~\bibnamefont {Hanada}},
  \bibinfo {author} {\bibfnamefont {S.}~\bibnamefont {Pateloudis}}, \bibinfo
  {author} {\bibfnamefont {E.}~\bibnamefont {Rinaldi}}, \bibinfo {author}
  {\bibfnamefont {A.}~\bibnamefont {Sch{\"a}fer}}, \bibinfo {author}
  {\bibfnamefont {P.}~\bibnamefont {Vranas}}, \ and\ \bibinfo {author}
  {\bibfnamefont {H.}~\bibnamefont {Watanabe}} (\bibinfo {collaboration} {Monte
  Carlo String/M-theory Collaboration}),\ }\bibfield  {title} {\enquote
  {\bibinfo {title} {{Confinement/deconfinement transition in the D0-brane
  matrix model \textemdash{} A signature of M-theory?}}}\ }\href {\doibase
  10.1007/JHEP05(2022)096} {\bibfield  {journal} {\bibinfo  {journal} {JHEP}\
  }\textbf {\bibinfo {volume} {2205}},\ \bibinfo {pages} {096} (\bibinfo {year}
  {2022}{\natexlab{a}})},\ \Eprint {http://arxiv.org/abs/2110.01312}
  {arXiv:2110.01312} \BibitemShut {NoStop}%
\bibitem [{\citenamefont {Schaich}\ \emph {et~al.}(2022)\citenamefont
  {Schaich}, \citenamefont {Jha},\ and\ \citenamefont
  {Joseph}}]{Schaich:2022duk}%
  \BibitemOpen
  \bibfield  {author} {\bibinfo {author} {\bibfnamefont {D.}~\bibnamefont
  {Schaich}}, \bibinfo {author} {\bibfnamefont {R.~G.}\ \bibnamefont {Jha}}, \
  and\ \bibinfo {author} {\bibfnamefont {A.}~\bibnamefont {Joseph}},\
  }\bibfield  {title} {\enquote {\bibinfo {title} {{Thermal phase structure of
  dimensionally reduced super-Yang--Mills}},}\ }\href {\doibase
  10.22323/1.396.0187} {\bibfield  {journal} {\bibinfo  {journal} {Proc. Sci.}\
  }\textbf {\bibinfo {volume} {LATTICE2021}},\ \bibinfo {pages} {187} (\bibinfo
  {year} {2022})},\ \Eprint {http://arxiv.org/abs/2201.03097}
  {arXiv:2201.03097} \BibitemShut {NoStop}%
\bibitem [{\citenamefont {Pateloudis}\ \emph {et~al.}(2022)\citenamefont
  {Pateloudis}, \citenamefont {Bergner}, \citenamefont {Bodendorfer},
  \citenamefont {Hanada}, \citenamefont {Rinaldi},\ and\ \citenamefont
  {Sch{\"a}fer}}]{Pateloudis:2022oos}%
  \BibitemOpen
  \bibfield  {author} {\bibinfo {author} {\bibfnamefont {S.}~\bibnamefont
  {Pateloudis}}, \bibinfo {author} {\bibfnamefont {G.}~\bibnamefont {Bergner}},
  \bibinfo {author} {\bibfnamefont {N.}~\bibnamefont {Bodendorfer}}, \bibinfo
  {author} {\bibfnamefont {M.}~\bibnamefont {Hanada}}, \bibinfo {author}
  {\bibfnamefont {E.}~\bibnamefont {Rinaldi}}, \ and\ \bibinfo {author}
  {\bibfnamefont {A.}~\bibnamefont {Sch{\"a}fer}},\ }\bibfield  {title}
  {\enquote {\bibinfo {title} {{Nonperturbative test of the Maldacena--Milekhin
  conjecture for the BMN matrix model}},}\ }\href {\doibase
  10.1007/JHEP08(2022)178} {\bibfield  {journal} {\bibinfo  {journal} {JHEP}\
  }\textbf {\bibinfo {volume} {2208}},\ \bibinfo {pages} {178} (\bibinfo {year}
  {2022})},\ \Eprint {http://arxiv.org/abs/2205.06098} {arXiv:2205.06098}
  \BibitemShut {NoStop}%
\bibitem [{\citenamefont {Pateloudis}\ \emph {et~al.}(2023)\citenamefont
  {Pateloudis}, \citenamefont {Bergner}, \citenamefont {Hanada}, \citenamefont
  {Rinaldi}, \citenamefont {Sch{\"a}fer}, \citenamefont {Vranas}, \citenamefont
  {Watanabe},\ and\ \citenamefont {Bodendorfer}}]{Pateloudis:2022ijr}%
  \BibitemOpen
  \bibfield  {author} {\bibinfo {author} {\bibfnamefont {S.}~\bibnamefont
  {Pateloudis}}, \bibinfo {author} {\bibfnamefont {G.}~\bibnamefont {Bergner}},
  \bibinfo {author} {\bibfnamefont {M.}~\bibnamefont {Hanada}}, \bibinfo
  {author} {\bibfnamefont {E.}~\bibnamefont {Rinaldi}}, \bibinfo {author}
  {\bibfnamefont {A.}~\bibnamefont {Sch{\"a}fer}}, \bibinfo {author}
  {\bibfnamefont {P.}~\bibnamefont {Vranas}}, \bibinfo {author} {\bibfnamefont
  {H.}~\bibnamefont {Watanabe}}, \ and\ \bibinfo {author} {\bibfnamefont
  {N.}~\bibnamefont {Bodendorfer}} (\bibinfo {collaboration} {Monte Carlo
  String/M-theory Collaboration}),\ }\bibfield  {title} {\enquote {\bibinfo
  {title} {{Precision test of gauge/gravity duality in D0-brane matrix model at
  low temperature}},}\ }\href {\doibase 10.1007/JHEP03(2023)071} {\bibfield
  {journal} {\bibinfo  {journal} {JHEP}\ }\textbf {\bibinfo {volume} {2303}},\
  \bibinfo {pages} {071} (\bibinfo {year} {2023})},\ \Eprint
  {http://arxiv.org/abs/2210.04881} {arXiv:2210.04881} \BibitemShut {NoStop}%
\bibitem [{\citenamefont {Catterall}\ \emph {et~al.}(2020)\citenamefont
  {Catterall}, \citenamefont {Giedt}, \citenamefont {Jha}, \citenamefont
  {Schaich},\ and\ \citenamefont {Wiseman}}]{Catterall:2020nmn}%
  \BibitemOpen
  \bibfield  {author} {\bibinfo {author} {\bibfnamefont {S.}~\bibnamefont
  {Catterall}}, \bibinfo {author} {\bibfnamefont {J.}~\bibnamefont {Giedt}},
  \bibinfo {author} {\bibfnamefont {R.~G.}\ \bibnamefont {Jha}}, \bibinfo
  {author} {\bibfnamefont {D.}~\bibnamefont {Schaich}}, \ and\ \bibinfo
  {author} {\bibfnamefont {T.}~\bibnamefont {Wiseman}},\ }\bibfield  {title}
  {\enquote {\bibinfo {title} {{Three-dimensional super-Yang--Mills theory on
  the lattice and dual black branes}},}\ }\href {\doibase
  10.1103/PhysRevD.102.106009} {\bibfield  {journal} {\bibinfo  {journal}
  {Phys. Rev. D}\ }\textbf {\bibinfo {volume} {102}},\ \bibinfo {pages}
  {106009} (\bibinfo {year} {2020})},\ \Eprint
  {http://arxiv.org/abs/2010.00026} {arXiv:2010.00026} \BibitemShut {NoStop}%
\bibitem [{\citenamefont {Sherletov}\ and\ \citenamefont
  {Schaich}(2022)}]{Sherletov:2022rnl}%
  \BibitemOpen
  \bibfield  {author} {\bibinfo {author} {\bibfnamefont {A.}~\bibnamefont
  {Sherletov}}\ and\ \bibinfo {author} {\bibfnamefont {D.}~\bibnamefont
  {Schaich}},\ }\bibfield  {title} {\enquote {\bibinfo {title} {{Investigations
  of supersymmetric Yang--Mills theories}},}\ }\href {\doibase
  10.22323/1.396.0031} {\bibfield  {journal} {\bibinfo  {journal} {Proc. Sci.}\
  }\textbf {\bibinfo {volume} {LATTICE2021}},\ \bibinfo {pages} {031} (\bibinfo
  {year} {2022})},\ \Eprint {http://arxiv.org/abs/2201.08626}
  {arXiv:2201.08626} \BibitemShut {NoStop}%
\bibitem [{\citenamefont {Sherletov}\ and\ \citenamefont
  {Schaich}(2023)}]{Sherletov:2023udh}%
  \BibitemOpen
  \bibfield  {author} {\bibinfo {author} {\bibfnamefont {A.}~\bibnamefont
  {Sherletov}}\ and\ \bibinfo {author} {\bibfnamefont {D.}~\bibnamefont
  {Schaich}},\ }\bibfield  {title} {\enquote {\bibinfo {title} {{Lattice
  Studies of 3D Maximally Supersymmetric Yang--Mills}},}\ }\href {\doibase
  10.22323/1.430.0221} {\bibfield  {journal} {\bibinfo  {journal} {Proc. Sci.}\
  }\textbf {\bibinfo {volume} {LATTICE2022}},\ \bibinfo {pages} {221} (\bibinfo
  {year} {2023})},\ \Eprint {http://arxiv.org/abs/2303.13880}
  {arXiv:2303.13880} \BibitemShut {NoStop}%
\bibitem [{\citenamefont {Schaich}\ and\ \citenamefont {Sherletov}(2025, in
  preparation)}]{Schaich:2024LAT}%
  \BibitemOpen
  \bibfield  {author} {\bibinfo {author} {\bibfnamefont {D.}~\bibnamefont
  {Schaich}}\ and\ \bibinfo {author} {\bibfnamefont {A.}~\bibnamefont
  {Sherletov}},\ }\bibfield  {title} {\enquote {\bibinfo {title} {{Maximally
  supersymmetric Yang--Mills in three dimensions}},}\ }\href
  {https://conference.ippp.dur.ac.uk/event/1265/contributions/6749/} {\bibfield
   {journal} {\bibinfo  {journal} {Proc. Sci.}\ }\textbf {\bibinfo {volume}
  {LATTICE2024}},\ \bibinfo {pages} {430} (\bibinfo {year} {2025, in
  preparation})}\BibitemShut {NoStop}%
\bibitem [{\citenamefont {Schaich}(2023)}]{Schaich:2022xgy}%
  \BibitemOpen
  \bibfield  {author} {\bibinfo {author} {\bibfnamefont {D.}~\bibnamefont
  {Schaich}},\ }\bibfield  {title} {\enquote {\bibinfo {title} {{Lattice
  studies of supersymmetric gauge theories}},}\ }\href {\doibase
  10.1140/epjs/s11734-022-00708-1} {\bibfield  {journal} {\bibinfo  {journal}
  {Eur. Phys. J. ST}\ }\textbf {\bibinfo {volume} {232}},\ \bibinfo {pages}
  {305--320} (\bibinfo {year} {2023})},\ \Eprint
  {http://arxiv.org/abs/2208.03580} {arXiv:2208.03580} \BibitemShut {NoStop}%
\bibitem [{\citenamefont {Golterman}\ and\ \citenamefont
  {Petcher}(1989)}]{Golterman:1988ta}%
  \BibitemOpen
  \bibfield  {author} {\bibinfo {author} {\bibfnamefont {M.~F.~L.}\
  \bibnamefont {Golterman}}\ and\ \bibinfo {author} {\bibfnamefont {D.~N.}\
  \bibnamefont {Petcher}},\ }\bibfield  {title} {\enquote {\bibinfo {title} {{A
  Local Interactive Lattice Model With Supersymmetry}},}\ }\href {\doibase
  10.1016/0550-3213(89)90080-1} {\bibfield  {journal} {\bibinfo  {journal}
  {Nucl. Phys. B}\ }\textbf {\bibinfo {volume} {319}},\ \bibinfo {pages}
  {307--341} (\bibinfo {year} {1989})}\BibitemShut {NoStop}%
\bibitem [{\citenamefont {Giedt}\ \emph {et~al.}(2004)\citenamefont {Giedt},
  \citenamefont {Koniuk}, \citenamefont {Poppitz},\ and\ \citenamefont
  {Yavin}}]{Giedt:2004vb}%
  \BibitemOpen
  \bibfield  {author} {\bibinfo {author} {\bibfnamefont {J.}~\bibnamefont
  {Giedt}}, \bibinfo {author} {\bibfnamefont {R.}~\bibnamefont {Koniuk}},
  \bibinfo {author} {\bibfnamefont {E.}~\bibnamefont {Poppitz}}, \ and\
  \bibinfo {author} {\bibfnamefont {T.}~\bibnamefont {Yavin}},\ }\bibfield
  {title} {\enquote {\bibinfo {title} {{Less naive about supersymmetric lattice
  quantum mechanics}},}\ }\href {\doibase 10.1088/1126-6708/2004/12/033}
  {\bibfield  {journal} {\bibinfo  {journal} {JHEP}\ }\textbf {\bibinfo
  {volume} {0412}},\ \bibinfo {pages} {033} (\bibinfo {year} {2004})},\ \Eprint
  {http://arxiv.org/abs/hep-lat/0410041} {hep-lat/0410041} \BibitemShut
  {NoStop}%
\bibitem [{\citenamefont {Bergner}\ \emph {et~al.}(2008)\citenamefont
  {Bergner}, \citenamefont {Kaestner}, \citenamefont {Uhlmann},\ and\
  \citenamefont {Wipf}}]{Bergner:2007pu}%
  \BibitemOpen
  \bibfield  {author} {\bibinfo {author} {\bibfnamefont {G.}~\bibnamefont
  {Bergner}}, \bibinfo {author} {\bibfnamefont {T.}~\bibnamefont {Kaestner}},
  \bibinfo {author} {\bibfnamefont {S.}~\bibnamefont {Uhlmann}}, \ and\
  \bibinfo {author} {\bibfnamefont {A.}~\bibnamefont {Wipf}},\ }\bibfield
  {title} {\enquote {\bibinfo {title} {{Low-dimensional Supersymmetric Lattice
  Models}},}\ }\href {\doibase 10.1016/j.aop.2007.06.010} {\bibfield  {journal}
  {\bibinfo  {journal} {Annals Phys.}\ }\textbf {\bibinfo {volume} {323}},\
  \bibinfo {pages} {946--988} (\bibinfo {year} {2008})},\ \Eprint
  {http://arxiv.org/abs/0705.2212} {arXiv:0705.2212} \BibitemShut {NoStop}%
\bibitem [{\citenamefont {Giedt}\ \emph {et~al.}(2018)\citenamefont {Giedt},
  \citenamefont {Lipstein},\ and\ \citenamefont {Martin}}]{Giedt:2018ygt}%
  \BibitemOpen
  \bibfield  {author} {\bibinfo {author} {\bibfnamefont {J.}~\bibnamefont
  {Giedt}}, \bibinfo {author} {\bibfnamefont {A.}~\bibnamefont {Lipstein}}, \
  and\ \bibinfo {author} {\bibfnamefont {P.}~\bibnamefont {Martin}},\
  }\bibfield  {title} {\enquote {\bibinfo {title} {{Lattice $\mathcal N = 4$
  three-dimensional super-Yang--Mills}},}\ }\href {\doibase
  10.22323/1.334.0239} {\bibfield  {journal} {\bibinfo  {journal} {Proc. Sci.}\
  }\textbf {\bibinfo {volume} {LATTICE2018}},\ \bibinfo {pages} {239} (\bibinfo
  {year} {2018})},\ \Eprint {http://arxiv.org/abs/1811.00516}
  {arXiv:1811.00516} \BibitemShut {NoStop}%
\bibitem [{\citenamefont {de~Wit}\ \emph {et~al.}(1988)\citenamefont {de~Wit},
  \citenamefont {Hoppe},\ and\ \citenamefont {Nicolai}}]{deWit:1988wri}%
  \BibitemOpen
  \bibfield  {author} {\bibinfo {author} {\bibfnamefont {B.}~\bibnamefont
  {de~Wit}}, \bibinfo {author} {\bibfnamefont {J.}~\bibnamefont {Hoppe}}, \
  and\ \bibinfo {author} {\bibfnamefont {H.}~\bibnamefont {Nicolai}},\
  }\bibfield  {title} {\enquote {\bibinfo {title} {{On the Quantum Mechanics of
  Supermembranes}},}\ }\href {\doibase 10.1016/0550-3213(88)90116-2} {\bibfield
   {journal} {\bibinfo  {journal} {Nucl. Phys. B}\ }\textbf {\bibinfo {volume}
  {305}},\ \bibinfo {pages} {545} (\bibinfo {year} {1988})}\BibitemShut
  {NoStop}%
\bibitem [{\citenamefont {Banks}\ \emph {et~al.}(1997)\citenamefont {Banks},
  \citenamefont {Fischler}, \citenamefont {Shenker},\ and\ \citenamefont
  {Susskind}}]{Banks:1996vh}%
  \BibitemOpen
  \bibfield  {author} {\bibinfo {author} {\bibfnamefont {T.}~\bibnamefont
  {Banks}}, \bibinfo {author} {\bibfnamefont {W.}~\bibnamefont {Fischler}},
  \bibinfo {author} {\bibfnamefont {S.~H.}\ \bibnamefont {Shenker}}, \ and\
  \bibinfo {author} {\bibfnamefont {L.}~\bibnamefont {Susskind}},\ }\bibfield
  {title} {\enquote {\bibinfo {title} {{M theory as a matrix model: A
  Conjecture}},}\ }\href {\doibase 10.1103/PhysRevD.55.5112} {\bibfield
  {journal} {\bibinfo  {journal} {Phys. Rev. D}\ }\textbf {\bibinfo {volume}
  {55}},\ \bibinfo {pages} {5112--5128} (\bibinfo {year} {1997})},\ \Eprint
  {http://arxiv.org/abs/hep-th/9610043} {hep-th/9610043} \BibitemShut {NoStop}%
\bibitem [{\citenamefont {Berenstein}\ \emph {et~al.}(2002)\citenamefont
  {Berenstein}, \citenamefont {Maldacena},\ and\ \citenamefont
  {Nastase}}]{Berenstein:2002jq}%
  \BibitemOpen
  \bibfield  {author} {\bibinfo {author} {\bibfnamefont {D.~E.}\ \bibnamefont
  {Berenstein}}, \bibinfo {author} {\bibfnamefont {J.~M.}\ \bibnamefont
  {Maldacena}}, \ and\ \bibinfo {author} {\bibfnamefont {H.~S.}\ \bibnamefont
  {Nastase}},\ }\bibfield  {title} {\enquote {\bibinfo {title} {{Strings in
  flat space and pp waves from $\mathcal N = 4$ super-Yang--Mills}},}\ }\href
  {\doibase 10.1088/1126-6708/2002/04/013} {\bibfield  {journal} {\bibinfo
  {journal} {JHEP}\ }\textbf {\bibinfo {volume} {0204}},\ \bibinfo {pages}
  {013} (\bibinfo {year} {2002})},\ \Eprint
  {http://arxiv.org/abs/hep-th/0202021} {hep-th/0202021} \BibitemShut {NoStop}%
\bibitem [{\citenamefont {Horowitz}\ and\ \citenamefont
  {Steif}(1990)}]{Horowitz:1989bv}%
  \BibitemOpen
  \bibfield  {author} {\bibinfo {author} {\bibfnamefont {G.~T.}\ \bibnamefont
  {Horowitz}}\ and\ \bibinfo {author} {\bibfnamefont {A.~R.}\ \bibnamefont
  {Steif}},\ }\bibfield  {title} {\enquote {\bibinfo {title} {{Space-Time
  Singularities in String Theory}},}\ }\href {\doibase
  10.1103/PhysRevLett.64.260} {\bibfield  {journal} {\bibinfo  {journal} {Phys.
  Rev. Lett.}\ }\textbf {\bibinfo {volume} {64}},\ \bibinfo {pages} {260}
  (\bibinfo {year} {1990})}\BibitemShut {NoStop}%
\bibitem [{\citenamefont {Kim}\ \emph {et~al.}(2003)\citenamefont {Kim},
  \citenamefont {Klose},\ and\ \citenamefont {Plefka}}]{Kim:2003rza}%
  \BibitemOpen
  \bibfield  {author} {\bibinfo {author} {\bibfnamefont {N.}~\bibnamefont
  {Kim}}, \bibinfo {author} {\bibfnamefont {T.}~\bibnamefont {Klose}}, \ and\
  \bibinfo {author} {\bibfnamefont {J.}~\bibnamefont {Plefka}},\ }\bibfield
  {title} {\enquote {\bibinfo {title} {{Plane wave matrix theory from $\mathcal
  N = 4$ super-Yang--Mills on $R \times S^3$}},}\ }\href {\doibase
  10.1016/j.nuclphysb.2003.08.019} {\bibfield  {journal} {\bibinfo  {journal}
  {Nucl. Phys. B}\ }\textbf {\bibinfo {volume} {671}},\ \bibinfo {pages}
  {359--382} (\bibinfo {year} {2003})},\ \Eprint
  {http://arxiv.org/abs/hep-th/0306054} {hep-th/0306054} \BibitemShut {NoStop}%
\bibitem [{\citenamefont {Lin}\ and\ \citenamefont
  {Maldacena}(2006)}]{Lin:2005nh}%
  \BibitemOpen
  \bibfield  {author} {\bibinfo {author} {\bibfnamefont {H.}~\bibnamefont
  {Lin}}\ and\ \bibinfo {author} {\bibfnamefont {J.~M.}\ \bibnamefont
  {Maldacena}},\ }\bibfield  {title} {\enquote {\bibinfo {title} {{Fivebranes
  from gauge theory}},}\ }\href {\doibase 10.1103/PhysRevD.74.084014}
  {\bibfield  {journal} {\bibinfo  {journal} {Phys. Rev. D}\ }\textbf {\bibinfo
  {volume} {74}},\ \bibinfo {pages} {084014} (\bibinfo {year} {2006})},\
  \Eprint {http://arxiv.org/abs/hep-th/0509235} {hep-th/0509235} \BibitemShut
  {NoStop}%
\bibitem [{\citenamefont {Ishii}\ \emph {et~al.}(2008)\citenamefont {Ishii},
  \citenamefont {Ishiki}, \citenamefont {Shimasaki},\ and\ \citenamefont
  {Tsuchiya}}]{Ishii:2008ib}%
  \BibitemOpen
  \bibfield  {author} {\bibinfo {author} {\bibfnamefont {T.}~\bibnamefont
  {Ishii}}, \bibinfo {author} {\bibfnamefont {G.}~\bibnamefont {Ishiki}},
  \bibinfo {author} {\bibfnamefont {S.}~\bibnamefont {Shimasaki}}, \ and\
  \bibinfo {author} {\bibfnamefont {A.}~\bibnamefont {Tsuchiya}},\ }\bibfield
  {title} {\enquote {\bibinfo {title} {{$\mathcal N = 4$ Super-Yang--Mills from
  the Plane Wave Matrix Model}},}\ }\href {\doibase 10.1103/PhysRevD.78.106001}
  {\bibfield  {journal} {\bibinfo  {journal} {Phys. Rev. D}\ }\textbf {\bibinfo
  {volume} {78}},\ \bibinfo {pages} {106001} (\bibinfo {year} {2008})},\
  \Eprint {http://arxiv.org/abs/0807.2352} {arXiv:0807.2352} \BibitemShut
  {NoStop}%
\bibitem [{\citenamefont {Ishiki}\ \emph {et~al.}(2009)\citenamefont {Ishiki},
  \citenamefont {Kim}, \citenamefont {Nishimura},\ and\ \citenamefont
  {Tsuchiya}}]{Ishiki:2009sg}%
  \BibitemOpen
  \bibfield  {author} {\bibinfo {author} {\bibfnamefont {G.}~\bibnamefont
  {Ishiki}}, \bibinfo {author} {\bibfnamefont {S.-W.}\ \bibnamefont {Kim}},
  \bibinfo {author} {\bibfnamefont {J.}~\bibnamefont {Nishimura}}, \ and\
  \bibinfo {author} {\bibfnamefont {A.}~\bibnamefont {Tsuchiya}},\ }\bibfield
  {title} {\enquote {\bibinfo {title} {{Testing a novel large-$N$ reduction for
  $\mathcal N = 4$ super-Yang--Mills theory on $R \times S^3$}},}\ }\href
  {\doibase 10.1088/1126-6708/2009/09/029} {\bibfield  {journal} {\bibinfo
  {journal} {JHEP}\ }\textbf {\bibinfo {volume} {0909}},\ \bibinfo {pages}
  {029} (\bibinfo {year} {2009})},\ \Eprint {http://arxiv.org/abs/0907.1488}
  {arXiv:0907.1488} \BibitemShut {NoStop}%
\bibitem [{\citenamefont {Ishiki}\ \emph {et~al.}(2011)\citenamefont {Ishiki},
  \citenamefont {Shimasaki},\ and\ \citenamefont {Tsuchiya}}]{Ishiki:2011ct}%
  \BibitemOpen
  \bibfield  {author} {\bibinfo {author} {\bibfnamefont {G.}~\bibnamefont
  {Ishiki}}, \bibinfo {author} {\bibfnamefont {S.}~\bibnamefont {Shimasaki}}, \
  and\ \bibinfo {author} {\bibfnamefont {A.}~\bibnamefont {Tsuchiya}},\
  }\bibfield  {title} {\enquote {\bibinfo {title} {{Perturbative tests for a
  large-$N$ reduced model of super-Yang--Mills theory}},}\ }\href {\doibase
  10.1007/JHEP11(2011)036} {\bibfield  {journal} {\bibinfo  {journal} {JHEP}\
  }\textbf {\bibinfo {volume} {1111}},\ \bibinfo {pages} {036} (\bibinfo {year}
  {2011})},\ \Eprint {http://arxiv.org/abs/1106.5590} {arXiv:1106.5590}
  \BibitemShut {NoStop}%
\bibitem [{\citenamefont {Costa}\ \emph {et~al.}(2015)\citenamefont {Costa},
  \citenamefont {Greenspan}, \citenamefont {Penedones},\ and\ \citenamefont
  {Santos}}]{Costa:2014wya}%
  \BibitemOpen
  \bibfield  {author} {\bibinfo {author} {\bibfnamefont {M.~S.}\ \bibnamefont
  {Costa}}, \bibinfo {author} {\bibfnamefont {L.}~\bibnamefont {Greenspan}},
  \bibinfo {author} {\bibfnamefont {J.}~\bibnamefont {Penedones}}, \ and\
  \bibinfo {author} {\bibfnamefont {J.}~\bibnamefont {Santos}},\ }\bibfield
  {title} {\enquote {\bibinfo {title} {{Thermodynamics of the BMN matrix model
  at strong coupling}},}\ }\href {\doibase 10.1007/JHEP03(2015)069} {\bibfield
  {journal} {\bibinfo  {journal} {JHEP}\ }\textbf {\bibinfo {volume} {1503}},\
  \bibinfo {pages} {069} (\bibinfo {year} {2015})},\ \Eprint
  {http://arxiv.org/abs/1411.5541} {arXiv:1411.5541} \BibitemShut {NoStop}%
\bibitem [{\citenamefont {Jha}\ \emph {et~al.}(2024)\citenamefont {Jha},
  \citenamefont {Joseph},\ and\ \citenamefont {Schaich}}]{data}%
  \BibitemOpen
  \bibfield  {author} {\bibinfo {author} {\bibfnamefont {R.~G.}\ \bibnamefont
  {Jha}}, \bibinfo {author} {\bibfnamefont {A.}~\bibnamefont {Joseph}}, \ and\
  \bibinfo {author} {\bibfnamefont {D.}~\bibnamefont {Schaich}},\ }\href@noop
  {} {\enquote {\bibinfo {title} {{Finite-temperature phase diagram of the BMN
  matrix model on the lattice --- data release}},}\ } (\bibinfo {year}
  {2024}),\ \bibinfo {note}
  {\href{https://doi.org/10.5281/zenodo.14391775}{doi:10.5281/zenodo.14391775}}\BibitemShut
  {NoStop}%
\bibitem [{\citenamefont {Furuuchi}\ \emph {et~al.}(2003)\citenamefont
  {Furuuchi}, \citenamefont {Schreiber},\ and\ \citenamefont
  {Semenoff}}]{Furuuchi:2003sy}%
  \BibitemOpen
  \bibfield  {author} {\bibinfo {author} {\bibfnamefont {K.}~\bibnamefont
  {Furuuchi}}, \bibinfo {author} {\bibfnamefont {E.}~\bibnamefont {Schreiber}},
  \ and\ \bibinfo {author} {\bibfnamefont {G.~W.}\ \bibnamefont {Semenoff}},\
  }\bibfield  {title} {\enquote {\bibinfo {title} {{Five-brane thermodynamics
  from the matrix model}},}\ }\href@noop {} {\  (\bibinfo {year} {2003})},\
  \Eprint {http://arxiv.org/abs/hep-th/0310286} {hep-th/0310286} \BibitemShut
  {NoStop}%
\bibitem [{\citenamefont {Spradlin}\ \emph {et~al.}(2004)\citenamefont
  {Spradlin}, \citenamefont {Van~Raamsdonk},\ and\ \citenamefont
  {Volovich}}]{Spradlin:2004sx}%
  \BibitemOpen
  \bibfield  {author} {\bibinfo {author} {\bibfnamefont {M.}~\bibnamefont
  {Spradlin}}, \bibinfo {author} {\bibfnamefont {M.}~\bibnamefont
  {Van~Raamsdonk}}, \ and\ \bibinfo {author} {\bibfnamefont {A.}~\bibnamefont
  {Volovich}},\ }\bibfield  {title} {\enquote {\bibinfo {title} {{Two-loop
  partition function in the planar plane-wave matrix model}},}\ }\href
  {\doibase 10.1016/j.physletb.2004.10.017} {\bibfield  {journal} {\bibinfo
  {journal} {Phys. Lett.}\ }\textbf {\bibinfo {volume} {B603}},\ \bibinfo
  {pages} {239--248} (\bibinfo {year} {2004})},\ \Eprint
  {http://arxiv.org/abs/hep-th/0409178} {hep-th/0409178} \BibitemShut {NoStop}%
\bibitem [{\citenamefont {Hadizadeh}\ \emph {et~al.}(2005)\citenamefont
  {Hadizadeh}, \citenamefont {Ramadanovic}, \citenamefont {Semenoff},\ and\
  \citenamefont {Young}}]{Hadizadeh:2004bf}%
  \BibitemOpen
  \bibfield  {author} {\bibinfo {author} {\bibfnamefont {S.}~\bibnamefont
  {Hadizadeh}}, \bibinfo {author} {\bibfnamefont {B.}~\bibnamefont
  {Ramadanovic}}, \bibinfo {author} {\bibfnamefont {G.~W.}\ \bibnamefont
  {Semenoff}}, \ and\ \bibinfo {author} {\bibfnamefont {D.}~\bibnamefont
  {Young}},\ }\bibfield  {title} {\enquote {\bibinfo {title} {{Free energy and
  phase transition of the matrix model on a plane-wave}},}\ }\href {\doibase
  10.1103/PhysRevD.71.065016} {\bibfield  {journal} {\bibinfo  {journal} {Phys.
  Rev. D}\ }\textbf {\bibinfo {volume} {71}},\ \bibinfo {pages} {065016}
  (\bibinfo {year} {2005})},\ \Eprint {http://arxiv.org/abs/hep-th/0409318}
  {hep-th/0409318} \BibitemShut {NoStop}%
\bibitem [{\citenamefont {Dhindsa}\ \emph {et~al.}(2022)\citenamefont
  {Dhindsa}, \citenamefont {Jha}, \citenamefont {Joseph}, \citenamefont
  {Samlodia},\ and\ \citenamefont {Schaich}}]{Dhindsa:2022uqn}%
  \BibitemOpen
  \bibfield  {author} {\bibinfo {author} {\bibfnamefont {N.~S.}\ \bibnamefont
  {Dhindsa}}, \bibinfo {author} {\bibfnamefont {R.~G.}\ \bibnamefont {Jha}},
  \bibinfo {author} {\bibfnamefont {A.}~\bibnamefont {Joseph}}, \bibinfo
  {author} {\bibfnamefont {A.}~\bibnamefont {Samlodia}}, \ and\ \bibinfo
  {author} {\bibfnamefont {D.}~\bibnamefont {Schaich}},\ }\bibfield  {title}
  {\enquote {\bibinfo {title} {{Non-perturbative phase structure of the bosonic
  BMN matrix model}},}\ }\href {\doibase 10.1007/JHEP05(2022)169} {\bibfield
  {journal} {\bibinfo  {journal} {JHEP}\ }\textbf {\bibinfo {volume} {2205}},\
  \bibinfo {pages} {169} (\bibinfo {year} {2022})},\ \Eprint
  {http://arxiv.org/abs/2201.08791} {arXiv:2201.08791} \BibitemShut {NoStop}%
\bibitem [{\citenamefont {Dhindsa}\ \emph {et~al.}(2024)\citenamefont
  {Dhindsa}, \citenamefont {Joseph}, \citenamefont {Samlodia},\ and\
  \citenamefont {Schaich}}]{Dhindsa:2023oes}%
  \BibitemOpen
  \bibfield  {author} {\bibinfo {author} {\bibfnamefont {N.~S.}\ \bibnamefont
  {Dhindsa}}, \bibinfo {author} {\bibfnamefont {A.}~\bibnamefont {Joseph}},
  \bibinfo {author} {\bibfnamefont {A.}~\bibnamefont {Samlodia}}, \ and\
  \bibinfo {author} {\bibfnamefont {D.}~\bibnamefont {Schaich}},\ }\bibfield
  {title} {\enquote {\bibinfo {title} {{Deconfinement Phase Transition in
  Bosonic BMN Model at General Coupling}},}\ }\href {\doibase
  10.1007/978-981-97-0289-3_274} {\bibfield  {journal} {\bibinfo  {journal}
  {Springer Proc. Phys.}\ }\textbf {\bibinfo {volume} {304}},\ \bibinfo {pages}
  {1020} (\bibinfo {year} {2024})},\ \Eprint {http://arxiv.org/abs/2308.02538}
  {arXiv:2308.02538} \BibitemShut {NoStop}%
\bibitem [{\citenamefont {Bergner}\ \emph
  {et~al.}(2022{\natexlab{b}})\citenamefont {Bergner}, \citenamefont
  {Catterall}, \citenamefont {Culver}, \citenamefont {Dhindsa}, \citenamefont
  {Giedt}, \citenamefont {Jha}, \citenamefont {Joseph}, \citenamefont
  {Schaich},\ and\ \citenamefont {Sherletov}}]{susy_code}%
  \BibitemOpen
  \bibfield  {author} {\bibinfo {author} {\bibfnamefont {G.}~\bibnamefont
  {Bergner}}, \bibinfo {author} {\bibfnamefont {S.}~\bibnamefont {Catterall}},
  \bibinfo {author} {\bibfnamefont {C.}~\bibnamefont {Culver}}, \bibinfo
  {author} {\bibfnamefont {N.~S.}\ \bibnamefont {Dhindsa}}, \bibinfo {author}
  {\bibfnamefont {J.}~\bibnamefont {Giedt}}, \bibinfo {author} {\bibfnamefont
  {R.~G.}\ \bibnamefont {Jha}}, \bibinfo {author} {\bibfnamefont
  {A.}~\bibnamefont {Joseph}}, \bibinfo {author} {\bibfnamefont
  {D.}~\bibnamefont {Schaich}}, \ and\ \bibinfo {author} {\bibfnamefont
  {A.}~\bibnamefont {Sherletov}},\ }\href@noop {} {\enquote {\bibinfo {title}
  {{SUSY LATTICE 2.3.1 --- Codes for supersymmetric lattice gauge theories}},}\
  } (\bibinfo {year} {2022}{\natexlab{b}}),\ \bibinfo {note}
  {\href{https://github.com/daschaich/susy}{github.com/daschaich/susy}}\BibitemShut
  {NoStop}%
\bibitem [{\citenamefont {Clark}\ and\ \citenamefont
  {Kennedy}(2007)}]{Clark:2006fx}%
  \BibitemOpen
  \bibfield  {author} {\bibinfo {author} {\bibfnamefont {M.~A.}\ \bibnamefont
  {Clark}}\ and\ \bibinfo {author} {\bibfnamefont {A.~D.}\ \bibnamefont
  {Kennedy}},\ }\bibfield  {title} {\enquote {\bibinfo {title} {{Accelerating
  dynamical fermion computations using the rational hybrid Monte Carlo (RHMC)
  algorithm with multiple pseudofermion fields}},}\ }\href {\doibase
  10.1103/PhysRevLett.98.051601} {\bibfield  {journal} {\bibinfo  {journal}
  {Phys. Rev. Lett.}\ }\textbf {\bibinfo {volume} {98}},\ \bibinfo {pages}
  {051601} (\bibinfo {year} {2007})},\ \Eprint
  {http://arxiv.org/abs/hep-lat/0608015} {hep-lat/0608015} \BibitemShut
  {NoStop}%
\bibitem [{\citenamefont {Schaich}\ and\ \citenamefont
  {DeGrand}(2015)}]{Schaich:2014pda}%
  \BibitemOpen
  \bibfield  {author} {\bibinfo {author} {\bibfnamefont {D.}~\bibnamefont
  {Schaich}}\ and\ \bibinfo {author} {\bibfnamefont {T.}~\bibnamefont
  {DeGrand}},\ }\bibfield  {title} {\enquote {\bibinfo {title} {{Parallel
  software for lattice $\mathcal N = 4$ supersymmetric Yang--Mills theory}},}\
  }\href {\doibase 10.1016/j.cpc.2014.12.025} {\bibfield  {journal} {\bibinfo
  {journal} {Comput. Phys. Commun.}\ }\textbf {\bibinfo {volume} {190}},\
  \bibinfo {pages} {200--212} (\bibinfo {year} {2015})},\ \Eprint
  {http://arxiv.org/abs/1410.6971} {arXiv:1410.6971} \BibitemShut {NoStop}%
\bibitem [{\citenamefont {Kuramashi}\ \emph {et~al.}(2020)\citenamefont
  {Kuramashi}, \citenamefont {Nakamura}, \citenamefont {Ohno},\ and\
  \citenamefont {Takeda}}]{Kuramashi:2020meg}%
  \BibitemOpen
  \bibfield  {author} {\bibinfo {author} {\bibfnamefont {Y.}~\bibnamefont
  {Kuramashi}}, \bibinfo {author} {\bibfnamefont {Y.}~\bibnamefont {Nakamura}},
  \bibinfo {author} {\bibfnamefont {H.}~\bibnamefont {Ohno}}, \ and\ \bibinfo
  {author} {\bibfnamefont {S.}~\bibnamefont {Takeda}},\ }\bibfield  {title}
  {\enquote {\bibinfo {title} {{Nature of the phase transition for finite
  temperature $N_f=3$ QCD with nonperturbatively $O(a)$ improved Wilson
  fermions at $N_t=12$}},}\ }\href {\doibase 10.1103/PhysRevD.101.054509}
  {\bibfield  {journal} {\bibinfo  {journal} {Phys. Rev. D}\ }\textbf {\bibinfo
  {volume} {101}},\ \bibinfo {pages} {054509} (\bibinfo {year} {2020})},\
  \Eprint {http://arxiv.org/abs/2001.04398} {arXiv:2001.04398} \BibitemShut
  {NoStop}%
\bibitem [{\citenamefont {Ferrenberg}\ and\ \citenamefont
  {Swendsen}(1988)}]{Ferrenberg:1988yz}%
  \BibitemOpen
  \bibfield  {author} {\bibinfo {author} {\bibfnamefont {A.~M.}\ \bibnamefont
  {Ferrenberg}}\ and\ \bibinfo {author} {\bibfnamefont {R.~H.}\ \bibnamefont
  {Swendsen}},\ }\bibfield  {title} {\enquote {\bibinfo {title} {{New Monte
  Carlo Technique for Studying Phase Transitions}},}\ }\href {\doibase
  10.1103/PhysRevLett.61.2635} {\bibfield  {journal} {\bibinfo  {journal}
  {Phys. Rev. Lett.}\ }\textbf {\bibinfo {volume} {61}},\ \bibinfo {pages}
  {2635--2638} (\bibinfo {year} {1988})}\BibitemShut {NoStop}%
\bibitem [{\citenamefont {Maldacena}(1998{\natexlab{b}})}]{Maldacena:1998im}%
  \BibitemOpen
  \bibfield  {author} {\bibinfo {author} {\bibfnamefont {J.~M.}\ \bibnamefont
  {Maldacena}},\ }\bibfield  {title} {\enquote {\bibinfo {title} {{Wilson loops
  in large-$N$ field theories}},}\ }\href {\doibase
  10.1103/PhysRevLett.80.4859} {\bibfield  {journal} {\bibinfo  {journal}
  {Phys. Rev. Lett.}\ }\textbf {\bibinfo {volume} {80}},\ \bibinfo {pages}
  {4859--4862} (\bibinfo {year} {1998}{\natexlab{b}})},\ \Eprint
  {http://arxiv.org/abs/hep-th/9803002} {hep-th/9803002} \BibitemShut {NoStop}%
\bibitem [{\citenamefont {Witten}(1998)}]{Witten:1998zw}%
  \BibitemOpen
  \bibfield  {author} {\bibinfo {author} {\bibfnamefont {E.}~\bibnamefont
  {Witten}},\ }\bibfield  {title} {\enquote {\bibinfo {title} {{Anti-de Sitter
  space, thermal phase transition, and confinement in gauge theories}},}\
  }\href {\doibase 10.4310/ATMP.1998.v2.n3.a3} {\bibfield  {journal} {\bibinfo
  {journal} {Adv. Theor. Math. Phys.}\ }\textbf {\bibinfo {volume} {2}},\
  \bibinfo {pages} {505--532} (\bibinfo {year} {1998})},\ \Eprint
  {http://arxiv.org/abs/hep-th/9803131} {hep-th/9803131} \BibitemShut {NoStop}%
\bibitem [{\citenamefont {Aharony}\ \emph
  {et~al.}(2004{\natexlab{a}})\citenamefont {Aharony}, \citenamefont {Marsano},
  \citenamefont {Minwalla},\ and\ \citenamefont {Wiseman}}]{Aharony:2004ig}%
  \BibitemOpen
  \bibfield  {author} {\bibinfo {author} {\bibfnamefont {O.}~\bibnamefont
  {Aharony}}, \bibinfo {author} {\bibfnamefont {J.}~\bibnamefont {Marsano}},
  \bibinfo {author} {\bibfnamefont {S.}~\bibnamefont {Minwalla}}, \ and\
  \bibinfo {author} {\bibfnamefont {T.}~\bibnamefont {Wiseman}},\ }\bibfield
  {title} {\enquote {\bibinfo {title} {{Black-hole--black-string phase
  transitions in thermal (1+1)-dimensional supersymmetric Yang--Mills theory on
  a circle}},}\ }\href {\doibase 10.1088/0264-9381/21/22/010} {\bibfield
  {journal} {\bibinfo  {journal} {Class. Quant. Grav.}\ }\textbf {\bibinfo
  {volume} {21}},\ \bibinfo {pages} {5169--5192} (\bibinfo {year}
  {2004}{\natexlab{a}})},\ \Eprint {http://arxiv.org/abs/hep-th/0406210}
  {hep-th/0406210} \BibitemShut {NoStop}%
\bibitem [{\citenamefont {Aharony}\ \emph
  {et~al.}(2004{\natexlab{b}})\citenamefont {Aharony}, \citenamefont {Marsano},
  \citenamefont {Minwalla}, \citenamefont {Papadodimas},\ and\ \citenamefont
  {Van~Raamsdonk}}]{Aharony:2003sx}%
  \BibitemOpen
  \bibfield  {author} {\bibinfo {author} {\bibfnamefont {O.}~\bibnamefont
  {Aharony}}, \bibinfo {author} {\bibfnamefont {J.}~\bibnamefont {Marsano}},
  \bibinfo {author} {\bibfnamefont {S.}~\bibnamefont {Minwalla}}, \bibinfo
  {author} {\bibfnamefont {K.}~\bibnamefont {Papadodimas}}, \ and\ \bibinfo
  {author} {\bibfnamefont {M.}~\bibnamefont {Van~Raamsdonk}},\ }\bibfield
  {title} {\enquote {\bibinfo {title} {{The Hagedorn/Deconfinement Phase
  Transition in Weakly Coupled Large-$N$ Gauge Theories}},}\ }\href {\doibase
  10.4310/ATMP.2004.v8.n4.a1} {\bibfield  {journal} {\bibinfo  {journal} {Adv.
  Theor. Math. Phys.}\ }\textbf {\bibinfo {volume} {8}},\ \bibinfo {pages}
  {603--696} (\bibinfo {year} {2004}{\natexlab{b}})},\ \Eprint
  {http://arxiv.org/abs/hep-th/0310285} {hep-th/0310285} \BibitemShut {NoStop}%
\bibitem [{\citenamefont {Imry}(1980)}]{Imry:1980zz}%
  \BibitemOpen
  \bibfield  {author} {\bibinfo {author} {\bibfnamefont {Y.}~\bibnamefont
  {Imry}},\ }\bibfield  {title} {\enquote {\bibinfo {title} {{Finite-size
  rounding of a first-order phase transition}},}\ }\href {\doibase
  10.1103/PhysRevB.21.2042} {\bibfield  {journal} {\bibinfo  {journal} {Phys.
  Rev. B}\ }\textbf {\bibinfo {volume} {21}},\ \bibinfo {pages} {2042--2043}
  (\bibinfo {year} {1980})}\BibitemShut {NoStop}%
\bibitem [{\citenamefont {Fisher}\ and\ \citenamefont
  {Berker}(1982)}]{Fisher:1982xt}%
  \BibitemOpen
  \bibfield  {author} {\bibinfo {author} {\bibfnamefont {M.~E.}\ \bibnamefont
  {Fisher}}\ and\ \bibinfo {author} {\bibfnamefont {A.~N.}\ \bibnamefont
  {Berker}},\ }\bibfield  {title} {\enquote {\bibinfo {title} {{Scaling for
  first-order transitions in thermodynamic and finite systems}},}\ }\href
  {\doibase 10.1103/PhysRevB.26.2507} {\bibfield  {journal} {\bibinfo
  {journal} {Phys. Rev. B}\ }\textbf {\bibinfo {volume} {26}},\ \bibinfo
  {pages} {2507--2513} (\bibinfo {year} {1982})}\BibitemShut {NoStop}%
\bibitem [{\citenamefont {Binder}\ and\ \citenamefont
  {Landau}(1984)}]{Binder:1984llk}%
  \BibitemOpen
  \bibfield  {author} {\bibinfo {author} {\bibfnamefont {K.}~\bibnamefont
  {Binder}}\ and\ \bibinfo {author} {\bibfnamefont {D.~P.}\ \bibnamefont
  {Landau}},\ }\bibfield  {title} {\enquote {\bibinfo {title} {{Finite-size
  scaling at first-order phase transitions}},}\ }\href {\doibase
  10.1103/PhysRevB.30.1477} {\bibfield  {journal} {\bibinfo  {journal} {Phys.
  Rev. B}\ }\textbf {\bibinfo {volume} {30}},\ \bibinfo {pages} {1477}
  (\bibinfo {year} {1984})}\BibitemShut {NoStop}%
\bibitem [{\citenamefont {Challa}\ \emph {et~al.}(1986)\citenamefont {Challa},
  \citenamefont {Landau},\ and\ \citenamefont {Binder}}]{Challa:1986sk}%
  \BibitemOpen
  \bibfield  {author} {\bibinfo {author} {\bibfnamefont {M.~S.~S.}\
  \bibnamefont {Challa}}, \bibinfo {author} {\bibfnamefont {D.~P.}\
  \bibnamefont {Landau}}, \ and\ \bibinfo {author} {\bibfnamefont
  {K.}~\bibnamefont {Binder}},\ }\bibfield  {title} {\enquote {\bibinfo {title}
  {{Finite size effects at temperature driven first order transitions}},}\
  }\href {\doibase 10.1103/PhysRevB.34.1841} {\bibfield  {journal} {\bibinfo
  {journal} {Phys. Rev. B}\ }\textbf {\bibinfo {volume} {34}},\ \bibinfo
  {pages} {1841--1852} (\bibinfo {year} {1986})}\BibitemShut {NoStop}%
\bibitem [{\citenamefont {Fukugita}\ \emph {et~al.}(1989)\citenamefont
  {Fukugita}, \citenamefont {Okawa},\ and\ \citenamefont
  {Ukawa}}]{Fukugita:1989yb}%
  \BibitemOpen
  \bibfield  {author} {\bibinfo {author} {\bibfnamefont {M.}~\bibnamefont
  {Fukugita}}, \bibinfo {author} {\bibfnamefont {M.}~\bibnamefont {Okawa}}, \
  and\ \bibinfo {author} {\bibfnamefont {A.}~\bibnamefont {Ukawa}},\ }\bibfield
   {title} {\enquote {\bibinfo {title} {{Order of the Deconfining Phase
  Transition in SU(3) Lattice Gauge Theory}},}\ }\href {\doibase
  10.1103/PhysRevLett.63.1768} {\bibfield  {journal} {\bibinfo  {journal}
  {Phys. Rev. Lett.}\ }\textbf {\bibinfo {volume} {63}},\ \bibinfo {pages}
  {1768} (\bibinfo {year} {1989})}\BibitemShut {NoStop}%
\bibitem [{\citenamefont {Stathopoulos}\ and\ \citenamefont
  {McCombs}(2010)}]{Stathopoulos:2010}%
  \BibitemOpen
  \bibfield  {author} {\bibinfo {author} {\bibfnamefont {A.}~\bibnamefont
  {Stathopoulos}}\ and\ \bibinfo {author} {\bibfnamefont {J.~R.}\ \bibnamefont
  {McCombs}},\ }\bibfield  {title} {\enquote {\bibinfo {title} {{PRIMME:
  preconditioned iterative multimethod eigensolver--methods and software
  description}},}\ }\href {\doibase 10.1145/1731022.1731031} {\bibfield
  {journal} {\bibinfo  {journal} {ACM Trans. Math. Softw.}\ }\textbf {\bibinfo
  {volume} {37}},\ \bibinfo {pages} {21} (\bibinfo {year} {2010})}\BibitemShut
  {NoStop}%
\bibitem [{\citenamefont {Foreman-Mackey}\ \emph {et~al.}(2013)\citenamefont
  {Foreman-Mackey}, \citenamefont {Hogg}, \citenamefont {Lang},\ and\
  \citenamefont {Goodman}}]{Foreman:2013mc}%
  \BibitemOpen
  \bibfield  {author} {\bibinfo {author} {\bibfnamefont {D.}~\bibnamefont
  {Foreman-Mackey}}, \bibinfo {author} {\bibfnamefont {D.~W.}\ \bibnamefont
  {Hogg}}, \bibinfo {author} {\bibfnamefont {D.}~\bibnamefont {Lang}}, \ and\
  \bibinfo {author} {\bibfnamefont {J.}~\bibnamefont {Goodman}},\ }\bibfield
  {title} {\enquote {\bibinfo {title} {{emcee: The MCMC Hammer}},}\ }\href
  {\doibase 10.1086/670067} {\bibfield  {journal} {\bibinfo  {journal} {PASP}\
  }\textbf {\bibinfo {volume} {125}},\ \bibinfo {pages} {306--312} (\bibinfo
  {year} {2013})},\ \Eprint {http://arxiv.org/abs/1202.3665} {arXiv:1202.3665}
  \BibitemShut {NoStop}%
\end{thebibliography}%
\end{document}